\documentclass[sigchi]{acmart}
\usepackage{xcolor}
\usepackage{xspace}
\usepackage{subcaption}
\definecolor{scarlet}{HTML}{fe4610}

\AtBeginDocument{%
  }

\copyrightyear{2026}
\acmYear{2026}
\setcopyright{cc}
\setcctype{by}
\acmConference[CHI '26]{Proceedings of the 2026 CHI Conference on Human Factors in Computing Systems}{April 13--17, 2026}{Barcelona, Spain}
\acmBooktitle{Proceedings of the 2026 CHI Conference on Human Factors in Computing Systems (CHI '26), April 13--17, 2026, Barcelona, Spain}
\acmPrice{}
\acmDOI{10.1145/3772318.3791843}
\acmISBN{979-8-4007-2278-3/2026/04}

\newcommand{\add}[1]{#1}
\newcommand{\delete}[1]{}

\newcommand{\final}[1]{#1}
\newcommand{\finaldelete}[1]{}

\begin{document}

\title[An Expert Schema for Evaluating LLM Errors]{An Expert Schema for Evaluating Large Language Model Errors in \\Scholarly Question-Answering Systems}

\author{Anna Martin-Boyle}
\affiliation{%
  \institution{University of Minnesota}
  \city{Minneapolis}
  \country{United States}}
\email{mart5877@umn.edu}

\author{William Humphreys}
\affiliation{%
  \institution{NASA Langley Research Center}
  \city{Hampton}
  \state{Virginia}
  \country{United States}}
\email{william.m.humphreys@nasa.gov}

\author{Martha Brown}
\affiliation{%
  \institution{NASA Langley Research Center}
  \city{Hampton}
  \state{Virginia}
  \country{United States}}
\email{martha.c.brown@nasa.gov}

\author{Cara Leckey}
\affiliation{%
  \institution{NASA Langley Research Center}
  \city{Hampton}
  \state{Virginia}
  \country{United States}}
\email{cara.ac.leckey@nasa.gov}

\author{Harmanpreet Kaur}
\affiliation{%
  \institution{University of Minnesota}
  \city{Minneapolis}
  \country{United States}}
\email{harmank@umn.edu}

\renewcommand{\shortauthors}{Martin-Boyle et al.}

\begin{abstract}
Large Language Models (LLMs) are transforming scholarly tasks like search and summarization, but their reliability remains uncertain. Current evaluation metrics for testing LLM reliability are primarily automated approaches that prioritize efficiency and scalability, but lack contextual nuance and fail to reflect how scientific domain experts assess LLM outputs in practice. We developed and validated a schema for evaluating LLM errors in scholarly question-answering systems that reflects the assessment strategies of practicing scientists. In collaboration with domain experts, we identified 20 error patterns across seven categories through thematic analysis of 68 question-answer pairs. We validated this schema through contextual inquiries with 10 additional scientists, which showed not only which errors experts naturally identify but also how structured evaluation schemas can help them detect previously overlooked issues. Domain experts use systematic assessment strategies, including technical precision testing, value-based evaluation, and meta-evaluation of their own practices. We discuss implications for supporting expert evaluation of LLM outputs, including opportunities for personalized, schema-driven tools that adapt to individual evaluation patterns and expertise levels. 

\end{abstract}



\maketitle

\section{Introduction}

Large Language Models (LLMs) are transforming how people interact with information for both everyday search, summarization, and writing; as well as information needs in complex domains like scientific research, mental health, law, and education. This transformation comes at a critical time: the volume of scientific literature is rapidly growing \cite{Bornmann2021,Krenn2023}, making it increasingly difficult for researchers to stay informed on recent developments~\cite{doi:10.1080/0194262X.2018.1445063,doi:10.1080/0194262X.2020.1758284,doi:10.1080/0194262X.2021.1991546}. 
These converging trends create urgent demand for reliable LLM-based QA systems to aid knowledge management~\cite{DNItrends}, particularly as these systems become integrated into search engines \cite{Venkatachary_2024} and digital libraries \cite{s2qa}. Yet as these systems become embedded in scholarly workflows, we lack understanding of how domain experts actually evaluate their outputs in practice, what errors they naturally identify versus miss, and what support they need for effective evaluation. 
\begin{figure*}[th!]
    \centering
    \includegraphics[width=\linewidth]{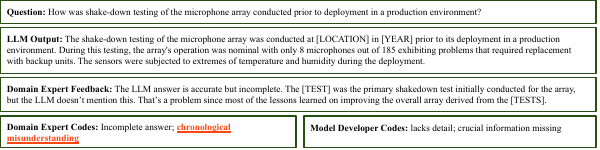}
    \Description{A worked example of the open coding process, presented as four labeled boxes stacked vertically, each outlined with a thin green border and separated by white space. The figure is primarily textual.

Box 1 is labeled ``Question:'' in bold and reads: ``How was shake-down testing of the microphone array conducted prior to deployment in a production environment?''

Box 2 is labeled ``LLM Output:'' in bold and reads: ``The shake-down testing of the microphone array was conducted at [LOCATION] in [YEAR] prior to its deployment in a production environment. During this testing, the array's operation was nominal with only 8 microphones out of 185 exhibiting problems that required replacement with backup units. The sensors were subjected to extremes of temperature and humidity during the deployment.'' Identifying details such as location and year have been replaced with bracketed placeholder tags for anonymity.

Box 3 is labeled ``Domain Expert Feedback:'' in bold and reads: ``The LLM answer is accurate but incomplete. The [TEST] was the primary shakedown test initially conducted for the array, but the LLM doesn't mention this. That's a problem since most of the lessons learned on improving the overall array derived from the [TESTS].'' Again, specific test names are replaced with bracketed placeholders.

Box 4 is split into two side-by-side cells. The left cell is labeled ``Domain Expert Codes:'' in bold and reads: ``Incomplete answer; chronological misunderstanding.'' The phrase ``chronological misunderstanding'' is visually highlighted with red text, distinguishing it from the plain-text code that precedes it. The right cell is labeled ``Model Developer Codes:'' in bold and reads: ``lacks detail; crucial information missing.''}
    \caption{Errors identified by domain experts and model developers with entity tags for anonymity. The expert recognized a \color{scarlet}\underline{\textbf{chronological error}} \color{black}about test sequences that the developer missed, showing how domain expertise can yield more precise error analysis.}
    \label{fig:example}
\end{figure*}

To reliably apply LLMs for scholarly QA, we must understand how they perform in domain-specific and information-intensive contexts. Current evaluation approaches for LLM-based QA systems primarily rely on automated metrics and benchmarks \cite{kamalloo-etal-2023-evaluating,xu-etal-2023-critical}. Although these offer efficient and scalable evaluations of LLM performance, prior work has shown that automated approaches do not always capture the domain-specific contextual nuances of the QA task \cite{krishna-etal-2021-hurdles}. Indeed, this aligns with broader critiques of automated benchmarking. \citet{NEURIPS2021Raji} argue that benchmarks often fail to capture the full complexity of real-world tasks, functioning more like arbitrary ``rooms'' in a museum than comprehensive evaluations. Work by other scholars similarly critiques how benchmarks lead to overfitting, gaming the system, and obscuring actual model limitations \cite{kiela-etal-2021-dynabench,bowman-dahl-2021-will}, and they often embed and perpetuate social biases~\cite{blodgett-etal-2021-stereotyping}. Together, these critiques show that benchmarks often become optimization targets rather than meaningful measures of progress, failing to predict real-world performance. Furthermore, these automated approaches are typically developed by model developers in isolation, missing the contextual nuances and evaluation strategies that domain experts bring to scholarly work.

We argue that automated approaches must be complemented with expert evaluations for knowledge management tasks such as scholarly QA. \add{Specifically, we focus on domain expert evaluation, where practicing scientists evaluate LLM outputs within their areas of expertise, rather than general end-user assessment. Domain experts possess the specialized knowledge required to identify subtle but critical errors that non-experts would miss. For example, a domain expert can recognize when causal relationships in experimental designs are misrepresented, when temporal sequences of technical developments are reversed (as shown in Figure 1), or when distinct theoretical frameworks are conflated despite superficial similarities.}
Such expert evaluations offer several \finaldelete{critical} advantages for scholarly QA systems by: (1) capturing experts' implicit quality criteria that they apply when evaluating outputs; (2) identifying errors that require deep domain knowledge to detect; and (3) evaluating genuine usefulness of LLM outputs for scholarly work, beyond factual (in)accuracies, to establish long-term value. 

In this work, we present an \add{expert-derived} schema of LLM errors for scholarly QA, developed and refined through a qualitative, contextual methodology \add{with practicing scientists evaluating outputs within their areas of research expertise.} Our approach brings together an interdisciplinary team of experts in identifying and categorizing these LLM errors across two phases.
In Phase 1, we systematically identified common error patterns in close collaboration with domain experts (N=3), who evaluated outputs from our scholarly QA system \add{on papers they had authored}. In Phase 2, we validated and refined our schema through a contextual inquiry and interview study with 10 additional domain experts from diverse science and engineering domains \add{on their authored papers}. Contextual inquiries capture domain expert evaluations of a system in real-time, as they conduct a task with the system under consideration. This data, coupled with detailed interviews, allowed us to capture an ecologically valid and contextual evaluation of LLMs for scholarly QA.\looseness=-1

Our two-phase approach resulted in an expert schema that comprises 20 error patterns organized into seven major categories, ranging from specificity on various types of hallucinations (e.g., fabricated citations, invented technical terms) to synthesis failures in multi-document contexts. Through \add{the second validation phase}, we found that while experts naturally identified errors in correctness and completeness, the structured schema \final{appeared to} help them detect previously overlooked issues, particularly subtle hallucinations and citation errors. Our qualitative analysis \finaldelete{revealed} \final{indicated} that experts employ \finaldelete{sophisticated} \final{systematic} evaluation strategies including technical precision testing and meta-evaluation of their own assessment practices. \add{We also analyzed the 188 questions experts posed across the two phases, identifying 11 question types and mapping error patterns to each, \finaldelete{revealing} \final{suggesting} that evaluation and mitigation strategies might be differentiated by question type.} Through this, we observed \finaldelete{significant} variation in which error types different experts prioritized, suggesting opportunities for personalized evaluation tools. 

We discuss the impact of our schema in serving as a foundation for developing \finaldelete{more} nuanced and ecologically valid quantitative benchmarks that \finaldelete{better} \final{are designed to} align with domain experts' evaluation priorities for scholarly QA. The schema offers structured evaluation (sub-)categories, some of which can be easily achieved via automated approaches while others might continue to require human-assisted evaluations. Our approach also offers insight into \finaldelete{and benefits of} how people differently evaluate LLMs in a contextual setting, as the schema categories add \finaldelete{significantly} more detail to the ideas behind automated benchmarks. We contextualize the high-level categories of our schema as guiding principles for detecting LLM errors, but also leave low-level instantiations of these errors to interpretive flexibility across domains, and end with a discussion of other ethical considerations. \looseness=-1

\section{Related Work}
\label{sec:related-work}

\subsection{Scholarly Question Answering}
\add{Scholarly question answering (QA) is a specialized NLP task that answers questions using scientific literature as the knowledge source. It} is a challenging NLP task due to the complexity, multimodality, and long context windows of scientific literature \cite{jin-etal-2019-pubmedqa,dasigi-etal-2021-dataset,Saikh2022,Auer2023,ruggeri-etal-2023-dataset,hilgert-etal-2024-evaluating,li-etal-2024-m3sciqa,pramanick2025spiqadatasetmultimodalquestion}. \add{In this setting, models must not only retrieve relevant passages, but also interpret and synthesize them into coherent answers that follow domain-specific conventions and constraints.}

\add{Several paradigms exist for implementing scholarly QA systems. Retrieval-augmented generation (RAG) approaches \cite{lewis2021retrievalaugmentedgenerationknowledgeintensivenlp} combine document retrieval with language model generation, first identifying relevant passages from a corpus and then generating answers conditioned on both the question and retrieved context. Knowledge graph-based approaches leverage structured representations of scientific knowledge, such as the Open Research Knowledge Graph (ORKG), which represents research contributions and enables comparison across scholarly articles \cite{Auer2023}. More recent systems employ LLMs directly on full-text scientific papers, though performance degrades with growing context length and varying positions of relevant information \cite{hilgert-etal-2024-evaluating}.}\looseness=-1

\add{The scholarly QA landscape has expanded} \finaldelete{significantly} with the integration of LLMs into search engines \cite{Venkatachary_2024} and digital libraries\add{, where commercial and research systems like Semantic Scholar now offer ``ask this paper'' functionality, enabling researchers to query individual papers or collections directly} \cite{s2qa}. However, even highly capable LLMs exhibit a well-documented tendency to produce plausible but incorrect answers \cite{bajpai-etal-2024-llms}, even in retrieval-augmented generation (RAG) systems \cite{munikoti-etal-2024-evaluating}. 
The growing volume of scientific literature, with publication rates increasing exponentially \cite{Bornmann2021,Krenn2023}, makes effective and trustworthy scholarly QA increasingly important for knowledge management. \add{Yet this same complexity makes scholarly QA particularly challenging to evaluate, as answers must satisfy not only factual accuracy but also domain-specific standards for precision, attribution, and contextual appropriateness.}

\subsection{Automated Evaluation of Question Answering}
\add{Automated evaluation approaches for QA systems have evolved from simple lexical matching to more sophisticated metrics, although \finaldelete{significant} limitations remain. Traditional metrics like exact match (EM) and F1 score remain widely used for extractive QA, measuring whether predicted answers match reference answers exactly or share overlapping tokens. BLEU \cite{papineni-etal-2002-bleu} and ROUGE \cite{lin-2004-rouge} scores, originally developed for machine translation and summarization respectively, have been adapted for QA evaluation by comparing generated answers to reference texts. However, these n-gram-based metrics limit the complexity of datasets that can be created and struggle with abstractive answers \cite{chen-etal-2019-evaluating}. Embedding-based metrics like BERTScore attempt to address this by comparing answers in learned semantic spaces, though \citet{chen-etal-2019-evaluating} show that both n-gram and embedding-based metrics correlate only moderately with human judgments on QA tasks. \citet{chen-etal-2020-mocha} introduce MOCHA and LERC, a learned metric trained on 40K human judgments that better approximates human ratings, while \citet{muttenthaler-etal-2020-unsupervised} exploit transformer internal representations to build an unsupervised correctness predictor for extractive QA.}

\add{Benchmark datasets provide standardized testbeds for comparing QA systems. For scholarly QA specifically, datasets like SciQA \cite{Auer2023}, QASA \cite{10.5555/3618408.3619195}, PubMedQA \cite{jin-etal-2019-pubmedqa}, ScienceQA \cite{Saikh2022}, and M3SciQA \cite{li-etal-2024-m3sciqa} provide expert-curated questions across scientific domains. These benchmarks enable systematic comparison but face critiques for potentially encouraging overfitting, gaming, and obscuring actual model limitations \cite{bowman-dahl-2021-will,kiela-etal-2021-dynabench}. In particular, \citet{NEURIPS2021Raji} argue that benchmarks often fail to capture the full complexity of real-world tasks, functioning more like arbitrary ``rooms'' in a museum than comprehensive evaluations. Furthermore, benchmarks often embed and perpetuate social biases \cite{blodgett-etal-2021-stereotyping} and become optimization targets rather than meaningful measures of progress.}

\add{These limitations become especially pronounced for long-form QA, where evaluation presents substantial difficulties \cite{krishna-etal-2021-hurdles,wang-etal-2024-revisiting}. \citet{xu-etal-2023-critical} demonstrate that no existing automatic metrics are predictive of human preference judgments, arguing for multi-faceted evaluation across dimensions such as factuality and completeness rather than single aggregate scores. \citet{otegi-etal-2020-automatic} show that exact-match and F1 metrics for QA over COVID-19 abstracts favor a model variant that \final{human annotators} actually dislike, while a lower-scoring variant is preferred in expert A/B tests. \citet{kamalloo-etal-2023-evaluating} find that lexical matching fails substantially when evaluating LLM-generated answers, and that automated evaluation models struggle to detect hallucinations---concluding that there appears to be no substitute for human evaluation. ProxyQA \cite{tan-etal-2024-proxyqa} proposes proxy-question-based evaluation and shows that it tracks majority human preference better than standard metrics. Across these efforts, errors are typically treated as undifferentiated correctness or factuality issues, rather than decomposed into domain-specific patterns. While automated approaches offer efficiency and scalability, they do not capture the domain-specific contextual nuances required for scholarly QA evaluation.}

\subsection{Human Evaluation of Question Answering}
\add{Human evaluation remains essential for assessing QA system quality, particularly for open-ended responses where multiple valid answers exist.} Evaluation approaches for QA systems include human evaluation, automated metrics, and hybrid methods \cite{celikyilmaz2021evaluationtextgenerationsurvey}. \add{Surveys such as \citet{10643089} organize open-domain QA evaluation into human, lexical, semantic, and LLM-based metrics.} Human evaluators better assess nuanced language comprehension \cite{10643089}, yet \citet{van-der-lee-etal-2019-best} found in their overview of human evaluation practices of automatically generated text that only 28\% of the studies they surveyed used domain experts.

Domain experts are essential for meaningful scholarly QA evaluation. Prior work has integrated domain experts into dataset creation and question development \cite{10.5555/3618408.3619195,peskoff-stewart-2023-credible,ruggeri-etal-2023-dataset,malaviya-etal-2024-expertqa}. For example, \citet{ruggeri-etal-2023-dataset} had experts engage in argumentative dialogues about their own papers to capture exploratory content beyond factuality. \citet{malaviya-etal-2024-expertqa} collected expert-curated questions across 32 fields and had the same experts evaluate LLM responses for attribution quality and factuality. Studies of expert evaluation practices identify sophisticated strategies beyond simple accuracy checks. Experts cross-reference claims against source materials, test for technical precision, and assess logical coherence \cite{peskoff-stewart-2023-credible}. Prior work identifies specific unreliability cues that experts naturally detect, including inconsistencies, hallucinated citations, and synthesis failures \cite{dasigi-etal-2021-dataset}. Theoretical work further supports the importance of domain expert involvement throughout NLP system development, with experts preferring human-guided approaches and bringing valuable perspectives on system design \cite{10.1145/3532106.3533483,10.1007/s10115-024-02212-5}.

The HCI community has increasingly recognized this evaluation crisis in LLM research and responded with human-centered approaches \cite{10.1145/3613905.3636302,10.1145/3706599.3706729}. Interactive evaluation tools have emerged as one response to these limitations. One approach focuses on assisting users in prompt optimization \cite{10.1145/3613904.3642016,10.1145/3613905.3650755,10.1145/3613904.3642216}. Others have taken a human-AI collaborative approach to evaluation \cite{10.1145/3613904.3641960,10.1145/3706598.3713181,10.1145/3706598.3713423}. Domain-specific evaluation frameworks have begun to emerge across various fields such as healthcare \cite{10.1145/3706599.3719675}, education \cite{10.1145/3613905.3650774,10.1145/3613904.3642414,10.1145/3706598.3713714}, and UX evaluation \cite{10.1145/3706598.3714045}. \add{SciEx uses human instructors to grade free-form exams and then shows that LLM-as-a-judge can approximate these grades without eliminating the need for expert oversight \cite{dinh-etal-2024-sciex}.}

\add{Recent frameworks also emphasize structured human evaluation of machine reading comprehension across multiple dimensions. \citet{schlegel-etal-2020-framework} propose evaluation datasets along axes such as linguistic complexity, required reasoning, background knowledge, factual correctness, and lexical cues, while \citet{sugawara-aizawa-2016-analysis} ground annotations in reading comprehension skills such as co-reference, logical inference, and causal reasoning. These capabilities are critical for scholarly QA, but they primarily describe what skills datasets require, rather than how errors manifest when experts interact with QA systems in context.} 

\citet{rodriguez-boyd-graber-2021-evaluation} identify two distinct evaluation paradigms derived from different research traditions: the Manchester paradigm, which probes system capabilities and understanding through expert evaluation, and the Cranfield paradigm, which prioritizes practical utility and end-user information needs via reusable test collections, or a hybrid between the two.
\add{Our work extends this human- and expert-centered line of research to RAG-style scholarly QA. Rather than treating experts solely as raters of overall quality or as calibrators for metric and benchmark design, we treat their in-situ assessments as first-class data for deriving a fine-grained error schema and for characterizing the evaluation strategies they employ when judging LLM outputs on their own papers.}\looseness=-1

\subsection{LLM Error Taxonomies}
\add{Beyond QA evaluation frameworks, there is now a rapidly growing body of work on hallucination and factuality in LLMs. Several comprehensive surveys synthesize definitions, taxonomies, and mitigation strategies at a systems level. \citet{10.1145/3703155} propose a taxonomy of hallucinations grounded in the LLM pipeline, distinguishing types by generation stage, knowledge source, and task setting, and review detection and mitigation methods as well as the limitations of retrieval-augmented LLMs for controlling hallucinations. \citet{10.1162/COLI.a.16} similarly survey hallucination phenomena across tasks, organizing prior work on detection, explanation, and mitigation, with an emphasis on how deviations from user input, prior context, or established world knowledge undermine reliability. \citet{10.1145/3742420} define the ``factuality issue'' as the probability that LLM outputs contradict established facts and analyze how LLMs store and retrieve factual knowledge, how factuality is evaluated through benchmarks and metrics, and how techniques such as retrieval augmentation and domain adaptation can reduce factual errors.\looseness=-1

Other work proposes more fine-grained hallucination taxonomies. \citet{rawte-etal-2023-troubling} define an extensive framework that profiles hallucinations along multiple axes, including degree (mild, moderate, alarming), orientation (factual mirage vs. silver lining), and category, and connect these distinctions to families of mitigation strategies. These efforts offer broad, task-agnostic characterizations of hallucination and factuality that are intended to apply across many LLM applications. By contrast, our work focuses narrowly on scholarly QA and retrieval-augmented assistants, deriving an error schema directly from domain experts' in-situ evaluations of answers about their own papers. While our categories overlap with some general notions of hallucination and factuality, they surface scholarly-specific failure modes---such as misattributed or fabricated citations, multi-paper synthesis failures, and subtle chronological or methodological misrepresentations---and tie them to the concrete evaluation practices that experts actually use. In this sense, our schema complements general hallucination and factuality surveys by providing a domain-specific instantiation grounded in expert behavior rather than system-level definitions alone.}

\section{Methods}
\add{In this section, we first describe our model setup for the scholarly QA task (Section~\ref{sec:system}), followed by our two-phase schema development approach with domain experts.} The first phase (Section \ref{sec:schema-dev-methods}) included schema generation in collaboration with a multi-disciplinary team of scientific domain experts (N=3) with between 15 and 40 years of research experience, a model developer (NLP specialist, N=1), and HCI researcher (N=1)---all authors of this paper. In Section~\ref{sec:schema-val-methods} we describe the second phase where we validated our schema with a different group of scientific domain experts (N=10). 

\add{Our choice of expert evaluation is grounded in the Manchester paradigm, one of two popular approaches for evaluating QA systems~\cite{rodriguez-boyd-graber-2021-evaluation}. The Manchester paradigm emphasizes probing system capabilities through expert assessment, since its primary goal is to test for human-like intelligence of the QA system under consideration. This paradigm is commonly used to create benchmark datasets for QA systems (e.g., QASA \cite{10.5555/3618408.3619195}). The other approach, the Cranfield paradigm, prioritizes the perspective of the end-user, thus seeking to evaluate whether the answer from a QA system meets the needs of the question-asker considering their ecological context. Our goals are more aligned with the Manchester paradigm---generating a schema of LLM errors that may be overlooked and may result in outcomes such as over-trust and misuse. That is, we ultimately seek to evaluate the intelligence of an LLM for scholarly QA. Therefore, we rely on expert paper authors in developing and validating our schema.}

\subsection{Scholarly Question-Answering \add{Setup}}\label{sec:system}
To evaluate expert assessment of LLM errors in scholarly QA, we first needed a concrete setup that would generate realistic outputs for experts to evaluate. We developed a retrieval-augmented generation (RAG) \cite{lewis2021retrievalaugmentedgenerationknowledgeintensivenlp} system designed to answer questions about scholarly papers while operating under constraints typical of many research settings. First, it reflects real-world scenarios where the technical solution must use an open-source model small enough to run on local infrastructure without external API calls. These requirements are driven by realistic security policies \add{(e.g., in safety-critical settings)} and budget limitations typical of many research institutions, and informed the technical decisions in our implementation. \add{Second, it follows recent work on small language models (SLMs), which argues that model size should be defined relative to task requirements and resource constraints rather than absolute parameter counts \cite{10.1145/3711896.3736563,10.1145/3768165}. 
SLMs are increasingly favored in practice for settings where privacy, local control over data, and predictable operating costs matter more than matching the absolute performance of frontier LLMs \cite{10.1145/3711896.3736563,10.1145/3768165}. Finally, our setup aligns with technical approaches that respond to growing concerns about the sustainability of large-scale AI systems, where recent work has highlighted the environmental and economic costs of deploying massive models \cite{10903774}.}\looseness=-1

\add{Critical to our evaluation goals, this choice of open-source infrastructure provided the interpretability necessary to distinguish between different error origins—whether failures arose from retrieval issues, model limitations, or semantic misunderstandings. With black-box commercial systems, we would be unable to determine whether an error stemmed from the retrieval component failing to find relevant passages, the model's parametric knowledge introducing hallucinations, or genuine reasoning failures. Our ability to inspect each component of the pipeline was essential for developing a comprehensive error taxonomy that could differentiate between these distinct failure modes. While our evaluation focused on a single open-source system, the resulting schema provides a foundation for systematic comparison across LLM architectures and scales.}
The following sections detail our task formulation, system architecture, and implementation choices.

\subsubsection{Task Description} We formalize our scholarly question-answering task as follows. Let $D=\{d_1,d_2,...,d_n\}$ represent a collection of scholarly documents, where each document $d_i$ contains text organized into sections. Given a natural language question $q$, our system must produce an answer $a$ that adheres to the following design objectives:
\begin{enumerate}
    \item The system is instructed to derive answers exclusively from content in $D$, with explicit attribution to source documents and paper sections. 
    \item The system comprises a retrieval-augmented generation pipeline, where the task decomposes into two stages:
    \begin{enumerate}
        \item Retrieval: $R(q,D)\rightarrow C$, where $C\subset D$ represents a context set of relevant passages selected based on semantic similarity to $q$.
        \item Generation: $G(q,C)\rightarrow a$, where the language model generates answer $a$ conditioned on both the question and retrieved context.
    \end{enumerate}
    \item Through prompt engineering, we instruct the model to maintain fidelity with the source material and avoid hallucination. The degree to which this is achieved varies and forms a key component of our evaluation.
    \item The system is prompted to include traceable citations for each claim.
\end{enumerate}

\noindent This formulation positions our work within closed-domain question answering, where answers should ideally be constrained to information present in a provided document collection. Unlike deterministic retrieval systems, our LLM-based approach introduces inherent uncertainty. The model may hallucinate information, misattribute sources, or fail to properly ground its responses despite explicit instructions. The retrieval-augmented approach aims to minimize these issues by providing relevant context, but perfect adherence to these objectives remains an open challenge in LLM-based systems \cite{munikoti-etal-2024-evaluating}.

\subsubsection{System Implementation} 
Our system implements the retrieval-augmented generation pipeline \cite{lewis2021retrievalaugmentedgenerationknowledgeintensivenlp} through several components, starting with a text preprocessing step, followed by information retrieval and then answer generation.

\subsubsection{Document Processing and Vectorization}
Each document in our collection $D$ is structured using a combination of PyMuPDF v1.23.5\footnote{\url{https://pypi.org/project/PyMuPDF/}, GNU Affero General Public License} and manual processing to ensure accuracy. Scholarly papers present preprocessing challenges such as complex multi-column layouts and embedded equations; our hybrid approach catches errors that automated parsers miss, such as incorrectly merged columns or omitted mathematical expression formatting. While a system deployed in the real world would need to handle this preprocessing programmatically, we did not want parsing errors to confound our evaluation of LLM-specific errors, as our focus is on characterizing failures in reasoning rather than artifacts of document extraction. We create sentence embeddings using NLTK v3.9.1\footnote{\url{https://github.com/nltk/nltk}, Apache 2.0}'s sentence tokenizer \cite{bird2009natural} and the Sentence Transformers \cite{reimers-gurevych-2019-sentence} model (All-MiniLM-L6-v2\footnote{\url{https://huggingface.co/sentence-transformers/all-MiniLM-L6-v2}, Apache-2.0}), stored in a vector-based index for semantic similarity search. 

\subsubsection{Retrieval} 
The retrieval process begins by encoding queries into the same vector space as the vector store. Let $Q_v$ denote the query vector and $S_v^i$ the vector representation of the $i$-th sentence in the corpus of $N$ sentences. We compute semantic similarity using cosine similarity:
\[
\text{Similarity}(Q_v,S_v^i)=\frac{Q_v\cdot S_v^i}{|Q_v||S_v^i|}
\]
The retrieval function $R(q,D)$ uses an iterative query expansion approach \cite{rocchio1971relevance,10.1145/2071389.2071390} to address several challenges in scholarly QA retrieval. Questions may use different terminology and go into less detail than source documents (e.g., researchers might ask about 'noise' when a paper discusses 'acoustic emissions', or request 'results' when the relevant content appears under 'experimental validation.') Additionally, initial retrieval may surface documents that use related technical concepts not present in the original question but are important for comprehensive retrieval. Query expansion addresses these gaps by incorporating domain-specific terminology discovered in the initial pass. In the first iteration, the system retrieves the top ($j=12$) sentences based on similarity scores. In multi-document settings, these retrievals are distributed evenly across documents. Initially, we encode query $q$ into the same vector space and retrieve the top $k$ most similar candidate sentences using cosine similarity. We iteratively expand the query by extracting keyphrases from the top candidates using KeyBERT \cite{grootendorst2020keybert}\footnote{\url{https://github.com/MaartenGr/KeyBERT}, MIT license}, appending the keyphrases to the original query, creating an expanded representation $q'$. The system continues retrieving sentences with $q'$ until reaching $n$ sentences to keep the final prompt within the model's 8K token limit. We set $k=12$ and $n=60$ based on iterative experimentation. While not exhaustively optimized, these values were effective in that they provided sufficient context without exceeding token limits across our evaluation questions.

\subsubsection{Context Formatting and Answer Generation}
The retrieved context $C$ is structured as JSON, where each sentence is annotated with its paper title, section header, and unique sentence ID. This metadata assists the model in attribution. The original query and retrieved sentences are concatenated with instructions for answering scholarly questions into a prompt, which is passed to Mixtral-8x-7B-Instruct-v0.1 (temperature=0.7)\footnote{\url{https://huggingface.co/mistralai/Mixtral-8x7B-Instruct-v0.1}, Apache-2.0} \cite{jiang2024mixtralexperts}. We ran the system via llama.cpp v0.2.61\footnote{\url{https://github.com/ggerganov/llama.cpp}, MIT} on four Tesla V100-PCIE-16GB GPUs \add{with 64GB total VRAM.} The full prompt is provided in Appendix \ref{app:prompt}.

\subsection{Phase 1: Schema Development Method}\label{sec:schema-dev-methods}
The schema development phase was completed by our multi- disciplinary research team, using the domain experts' knowledge to generate QA pairs, and adding everyone's perspectives in qualitatively coding the answers for potential errors. \add{The domain experts came from science and engineering fields (Physics and Physical Chemistry, Materials Science); with between 15 and 40 years of research experience; and varied experience with LLMs (1 used them for scholarly and general-purpose QA, 1 only for scholarly tasks, and 1 infrequently though aware of capabilities).}  

First, the two domain experts wrote questions about papers they had authored \add{based on instructions to compose both single- and multi-document questions. We instructed them that these should be factual, analytical, and synthesizing questions across a variety of aspects, including but not limited to positioning within related work, methodological choices, interpretation of results, and implications of findings}. These questions probed the system's ability to extract, interpret, and synthesize information from single and multiple papers. \add{The inclusion criteria for paper selection were that the articles were authored by the participating expert, were publicly available, and were sufficiently related that meaningful multi-document questions could be posed across them.} One expert wrote 25 single- and three multi-document questions across five papers \add{in the aeroacoustics domain}, and the other wrote 32 single- and eight multi-document questions across four papers \add{in the non-destructive evaluation domain}; giving us a total sample of 68 QA pairs. The experts then wrote feedback for each response they received from our scholarly QA system. We relied on two, instead of all three domain experts on the research team, for this QA pair generation to preserve one expert perspective purely for qualitative coding.\looseness=-1

Second, we conducted inductive thematic analysis following \citet{Braun01012006}'s approach of open and axial coding. We first open coded each line of expert feedback for all QA pairs (i.e., capturing the essence of the feedback in a descriptive way), which was done by the two experts who generated the QA pairs and the model developer, as people with closest expertise about the content and system. To facilitate this challenging and open-ended task of evaluating LLM outputs on the part of domain experts (i.e., team members with no expertise in ML or HCI), we provided sensitizing concepts in the form of ``errors of inclusion'' (factually incorrect/irrelevant information) and ``errors of omission'' (correctness, relevance, and completeness) as common criteria for human evaluation of QA outputs \cite{10643089}. Experts were encouraged to elaborate their observations beyond broad categories. Each expert coded their own QA pairs and the NLP specialist coded all 68 samples. Finally, the whole research team met over three two-hour collaborative sessions to conduct axial coding of the open codes, i.e., the process of identifying common axes in the descriptive open code data, which resulted in the initial schema.  

\subsection{Phase 2: Schema Validation Method}\label{sec:schema-val-methods}
We validated our schema with 10 additional domain experts \add{(2 women and 8 men)} from science and engineering fields including aeroacoustics, aeroscience, autonomous systems, computer engineering, materials science, mechanical engineering, safety critical systems, and systems analysis, \add{with 10--45 years of research experience. These experts had varied familiarity with LLMs: three regularly used LLMs for scholarly tasks; four used them for general-purpose QA; and three used them infrequently, only for maintaining awareness of a new technology.} We recruited them through internal networks at NASA Langley Research Center. Participation was voluntary and uncompensated. All experts were informed about the study's purpose. Our institution's IRB determined this work was exempt from review.

First, each expert selected three English-language papers \add{they had authored} and wrote twelve questions: three single-document questions for each paper and three multi-document questions covering all papers. The experts were informed they could ``write probing questions that are factual (testing the LLM's ability to extract specific information from the papers), analytical (requiring the LLM to interpret or analyze information from the papers), or synthesizing (asking the LLM to combine information from different parts of the paper or from multiple papers).''

Second, we conducted an evaluation of the LLM answers to expert questions via a think aloud contextual inquiry~\cite{10.1145/203356.203365} and an interview study: experts discussed their reactions to the LLM answers out loud, as they were reading them for the first time. Prior to each expert feedback session, we used our system to generate responses to each expert's questions. We conducted two-hour feedback sessions with each expert (N=10), which were recorded and transcribed to facilitate analysis. These sessions were structured in two parts: in part one, experts freely noted shortcomings in LLM answers to their questions without being primed by our schema; and in part two, experts applied an inventory of questions we wrote to operationalize our schema (Appendix \ref{app:inventory}). The inventory did not include system failure errors, as the accurate identification of these errors requires LLM knowledge which we did not require in our experts. We did not require our experts to have this knowledge, and so eliminated the error type to avoid confusion.

This study design allowed us to see which errors emerged organically and if they were similar to our schema, and then observe if our schema could help identify errors that were missed. We followed the same inductive thematic analysis approach as before, consolidating the open codes from the experts' freeform feedback using our schema codes when applicable and creating new ones as needed; new codes are highlighted in yellow in Tables \ref{tab:schema-1} and \ref{tab:schema-2}. \add{In this qualitative coding process, we also captured two procedural aspects of schema validation and use: (1) patterns in experts' assessment processes and priorities; and (2) the types of questions they asked, along with their rationale for picking these questions for evaluation. We describe these in our Results along with the final schema.}\looseness=-1

\section{Results}
\add{We first describe our final expert-driven schema of LLM errors, collating results from both phases of schema development and validation. Next, we present themes of experts' evaluation patterns, and finally a categorization of the questions they asked mapped to the schema's error categories.}

\subsection{\add{Expert Schema of LLM Errors}}\label{sec:schema-dev-findings}
The schema is derived from open and axial coding of LLM errors identified in \textbf{Phase 1} (schema development with the multi-disciplinary research team) and \textbf{Phase 2} (evaluation with domain experts). In \textbf{Phase 1}, the open coding resulted in 49 unique error codes: 20 overlapping between the stakeholders, 14 unique to the developer, and 15 unique to the experts (see Table \ref{tab:open-codes} in Appendix \ref{app:open-codes}). The developer focused on NLP issues such as the model's ability to extract information from text and process different information formats. In contrast, the domain experts identified codes that emphasized scholarly quality and consistency. They also considered the potential impact of the error on a human reader as indicated by the term ``misleading,'' and \finaldelete{were more likely to identify} \final{considered} positive attributes, noting qualities like  ``complete'', and ``accurate'', while the model developer had a neutral code called ``no issues''. These codes about misleading content were ultimately removed during axial coding, as misleading responses could stem from many error types in our schema; ``misleading'' is more a consequence of an error than an error itself.
\begin{table*}[ht!]

    \centering
    \begin{tabular}{p{.015\textwidth}p{.75\textwidth}p{.05\textwidth}p{.05\textwidth}} \toprule
        \textbf{ID} & \textbf{Title \& Description} & Part 1 & Part 2 \\\midrule
        \midrule
         \multicolumn{4}{c}{\textbf{Incorrect Answer}} \\\midrule 
         1 & \textbf{Completely incorrect:} Answer is entirely wrong. & 6 & 0 \\
         2 & \textbf{Partially incorrect:} Mix of correct and incorrect information.  & - & - \\
         2.a & \textbf{Basic accuracy issues:} High-level concepts correct but details wrong.  & 8 & 2 \\
         2.b & \textbf{Misinterpretation:} Incorrect understanding of paper content.  & 10 & 0 \\
         2.c & \textbf{Self-contradiction:} Contains internal logic errors or contradictions.  & 5 & 0 \\
         2.d & \textbf{Reference confusion:} Mixes up sources and their content.  & 4 & 0 \\
         2.e & \textbf{Multimedia interpretation:} Incorrect interpretation of tables, equations, etc.  & 3 & 0 \\
         3 & \textbf{Contains hallucinations:} Generated information appears plausible but is factually incorrect or completely fabricated.  & 6 & 4\\
         4 & \textbf{Superfluous content:} Includes irrelevant or unnecessary information.  & 8 & 1 \\ 
         \midrule
         \multicolumn{4}{c}{\textbf{Contains Hallucinations}} \\\midrule
         3.a & \textbf{Terminology:} inventing or wrongly defining terms and acronyms. & 1 & 0 \\
         3.b & \textbf{Numerical data:} Making up statistics, measurements, performance metrics, quantitative results, and other numerical information & 1 & 0\\
         3.c & \large$\ast$ \normalsize\textbf{Mathematical constructs:} Inventing equations, formulae, and mathematical relationships & 2 & - \\
         3.d & \large$\ast$ \normalsize\textbf{Technical content:} False technical characteristics or behaviors, incorrect system properties, invented technical limitations or capabilities, and any other made up information internal to the paper. & 3 & - \\
         3.e & \large$\ast$ \normalsize\textbf{External context and world knowledge:} General scientific knowledge and related literature. & 3 & - \\
         3.f & \textbf{Document structure:} Inventing page, section, table, and figure numbers. & 1 & 2 \\
         3.g & \textbf{Citation information:} Fabricating entire citations, or parts of citations such as author names, dates, titles, publication venues, and DOI and page numbers. & 1 & 8 \\
         \bottomrule
    \end{tabular}
    \caption{Typology of incorrect answers and hallucinations found in LLM responses to scholarly questions. The columns Part 1 and Part 2 refer to the number of experts who identified each error during unprimed evaluation and structured evaluation of the validation study, respectively. Asterisks ($\ast$) indicate additional error types discovered during the schema validation phase. Dashes in column Part 2 indicate error types that were not included in our inventory, as these errors were not identified during the schema development process.  Examples are in Appendix \ref{sec:examples}. }
    \label{tab:schema-1}
\end{table*}

\begin{table*}[ht!]
    \centering
    \begin{tabular}{p{.015\textwidth}p{.75\textwidth}p{.05\textwidth}p{.05\textwidth}} \toprule
        \textbf{ID} & \textbf{Title \& Description} & \textbf{Part 1} & \textbf{Part 2} \\\midrule\midrule
         \multicolumn{4}{c}{\textbf{Incomplete Answer}}  \\\midrule
         5 & \textbf{Major omissions:} Important content or sections are left out of the answer. & - & -\\
         5.a & \textbf{Ignored whole section:} Entire relevant sections of content are overlooked.  & 4 & 1 \\
         5.b & \textbf{Missed main point:} Core argument or central thesis is not captured.  &  8 & 1 \\
         5.c & \textbf{Ignored whole paper:} Failed to consider one or more papers in multi-document scenarios.  & 1 & 2 \\
         5.d & \textbf{Incomplete definition:} Definition lacks crucial information.  & 6 & 1 \\
         5.e & \textbf{Incomplete references:} Missing important citations or source materials.  & 2 & 3 \\
         5.f & \textbf{Chronological gaps:} Timeline missing key events or developments.  & 0 & 1\\
         6 & \textbf{Lacking details:} Higher-level concepts present but missing lower-level details.  & 8 & 1\\
         7 & \textbf{Multimedia comprehension:} Failed to connect different information formats effectively, for example not linking table contents with its caption and in-text references.  & 4 & 0 \\
         
         \midrule
         \multicolumn{4}{c}{\textbf{Question Interpretation}} \\\midrule
         8 & \textbf{Question misinterpretation:} Wrong understanding leading to incorrect answer.  & 8 & 1 \\
         9 & \textbf{Question redirection:} Creates and answers a different question.  & 1 & 2 \\
         \midrule
         \multicolumn{4}{c}{\textbf{Synthesis Issues}} \\\midrule
         10 & \textbf{Document linking failure:} Unable to connect information to the correct document in a multi-document setting.  & 2 & 1 \\
         11 & \textbf{Chronological confusion:} Misunderstanding temporal relationships.  & 0 & 0 \\
         12 & \textbf{Disjointed response:} Fragmented or poorly integrated answer. For example, each paper is discussed in isolation.  & 8 & 1\\
         13 & \textbf{Source confusion:} Incorrectly attributes information across papers.  & 3 & 2 \\
         \midrule
         \multicolumn{4}{c}{\textbf{Formatting Issues}} \\\midrule
         14 & \textbf{Verbosity:} Unnecessarily wordy or repetitive.  & 8 & 0 \\
         15 & \textbf{Notation errors:} Incorrect use of technical notation.  & 1 & 1\\
         16 & \textbf{Language issues:} Improper wording, spelling, grammar, or syntax.  & 5 & 0 \\
         17 & \textbf{Inconsistent referencing:} Varies in how content is referenced.  & 1 & 0 \\
         \midrule
         \multicolumn{4}{c}{\textbf{System Failures}} \\\midrule
         18 & \textbf{LLM limitations:} Issues inherent to the model being used, such as limited context length, lack of multi-modal understanding, and inability to access external domain knowledge and digital libraries.  & 4 & - \\
         19 & \textbf{Retrieval failures:} Required context not provided by the retriever.  & 0 & - \\
         20 & \textbf{Data structure issues:} Problems with internal data organization leaking into system outputs.  & 0 & - \\ 
         \bottomrule
    \end{tabular}
    \caption{Additional error categories and system failures identified in LLM responses to scholarly questions. 
    Column Part 2 does not have values for \textbf{system failures} because we did not include questions in the inventory that would address these errors.\looseness=-1}
    \label{tab:schema-2}
\end{table*}

The collaborative axial coding sessions allowed us to consolidate open codes into the seven top-level categories shown in Tables~\ref{tab:schema-1} and \ref{tab:schema-2}. In \textbf{Phase 2}, all seven categories surfaced again in inductive coding of experts' unprimed feedback, \finaldelete{showing} \final{suggesting} that expert feedback on LLM use for scholarly QA may be consistent. Note that this consistency did not come from us applying our axial codes to confirm the use of our schema; rather, the same axial codes emerged from an inductive coding of this new data. Some new open codes emerged under the same axial categories; these \textbf{Phase 2}-only additions are highlighted in yellow in Tables~\ref{tab:schema-1} and \ref{tab:schema-2}.

Below, we summarize each category and its key sub-types, drawing on evidence from both phases. Codes are not mutually exclusive; when an answer exhibited multiple failure modes (e.g., omissions plus factual errors), we applied all relevant codes.

\subsubsection{\add{Incorrect Answers}}
\add{From \textbf{Phase 1}, we coded \textit{Incorrect Answers}} as ranging from \textit{completely incorrect} to specific types of \textit{partially incorrect}. Inaccuracies were commonly \textit{completely incorrect} when the model's answer failed to capture the paper's central claims. For example, when comparing ``older versus newer'' studies, the model offered generic contrasts and missed the central methodological advance described by the paper, resulting in a substantively incorrect description of what actually changed. \textit{Partially incorrect} answers ranged from \textit{basic accuracy issues}, where high-level concepts were correct but details like parameters, values, or terminology were wrong, to subtle \textit{misinterpretation} of the main paper content or incorrect \textit{multimedia interpretation} of content such as figures, tables, and equations. For instance, when asked about specific test results, the system incorrectly stated that ``percent differences between EFIT and dispersion curve group velocities are less than 2\%, while the percent differences between EFIT and experiment are as large as 16\%'' when the reference indicated differences ranging from 1-14\%. In explaining a method used by the author, the answer mixed up notation and equations, using ``$p_x$'' where the paper uses ``$p_a(x,t)$'', and ``$p_y$'' where it uses ``$p_s(y,r')$.'' The surrounding text was broadly accurate, but the equation mistakes made the result formally incorrect. We also observed \textit{self-contradiction}, where responses contained internal inconsistencies (e.g., asserting a claim and later reversing it or reporting mutually incompatible values), making the output unreliable even when individual sentences sounded plausible. Finally, some incorrect answers included \textit{superfluous content}, adding tangential background or speculative commentary that obscured the core error and made verification more difficult.

\add{In \textbf{Phase 2}, at least one sub-type of \textit{Incorrect Answers} came up in all 10 experts' unprimed evaluation, with several sub-types being recognized by a majority.} Many experts such as Expert 3 (E3) identified cases where the LLM was wrong because it misunderstood basic scientific principles or because it conflated technical concepts (E3, E7, E9). E1, E6, and E8-10 were concerned with plausible-sounding but incorrect statements which would be misleading to non-experts. \add{This is evident in the high number of unprimed mentions of the \textit{partially incorrect} sub-type of \textit{Incorrect Answers} (see column Part 1 in Table \ref{tab:schema-1}). A primed evaluation using our schema (see column Part 2) did not yield \final{notably} more \textit{Incorrect Answers} being recognized, likely an artifact of the initially high number in Part 1.}

\subsubsection{\add{Hallucinations}}
\add{From our \textbf{Phase 1} analysis, \textit{Hallucinations}} emerged as a \finaldelete{distinct} \final{notable} category due to their frequency and severity. \add{This included \textit{terminology} as a common sub-type:} in one case, when asked about ``BVI noise,'' the system incorrectly expanded the acronym as ``Boundary Value Problem'' rather than ``Blade Vortex Interaction,'' then proceeded to discuss monopole terms when BVI actually relates to dipole terms, which shows how hallucinations can propagate through the response. The LLM created internally consistent but incorrect explanations that might appear credible to non-experts.
\add{Hallucinations of \textit{citation information} were pervasive in \textbf{Phase 1}, appearing across multi-document contexts. The system frequently fabricated components of otherwise-valid references, including incorrect DOIs, journal names, and erroneous author information. It also hallucinated aspects of \textit{document structure}, such as non-existent page numbers (e.g., referencing page ``13'' in a 10-page article) or tables that did not exist (table ``19'' when only 4 tables were present). In one case, the system produced a reference where ``other than [\textsc{Author Name}] all other author information is not accurate.''} We also observed hallucinations of \textit{numerical data}, where the system invented quantitative results such as performance metrics or percentage differences that were not supported by the source documents.

\add{In \textbf{Phase 2}, unprimed mentions of specific types of hallucinations were surprisingly low, with only 1-3 (out of 10) experts mentioning them. Despite considerable focus on hallucinated \textit{citation information} in conversations about LLM errors in general media, even this sub-type was rarely caught,} only by chance as when E6 noticed errors in their own name. Indeed, specific sub-types of hallucinations were \final{among the most commonly identified categories} of errors discovered using our schema-based inventory during the primed \textbf{Phase 2} evaluation. With the inventory guiding them on specific types of potential hallucinations, experts were able to identify hallucinated \textit{citation information} (E2-5 and E7). E1-2 and E6 noted additional instances of missing source materials such as tables and figures. They also found new cases with incorrectly attributed information within (E10) and across (E2) papers. 

Thinking about hallucinations with the schema priming them also \final{appeared to} help experts identify other types of issues using the inventory, including \textit{missed main point} (E6), \textit{chronological gaps} (E10), \textit{disjointed response} (E9), \textit{basic accuracy issues} (E1, E10), and \textit{Question Interpretation} errors (E5-6). \add{Importantly, \textbf{Phase 2} also surfaced several hallucination sub-types that were not identified during \textbf{Phase 1} open coding and were therefore added to the final schema (highlighted in yellow in Table \ref{tab:schema-1}), including \textit{mathematical constructs}, \textit{technical content}, and \textit{external context and world knowledge}. \add{One salient example was \textit{Hallucinations} of \textit{technical content} where the system generated plausible-sounding but fundamentally incorrect explanations. When asked ``What is a diffractive optic?'', the system responded that it ``modifies the direction, phase, amplitude, or polarization of light through processes such as reflection, refraction, absorption, and diffusion.'' E10 noted: ``The diffractive optics work based on the principle of diffraction as in what's in their name. Reflection, refraction, absorption and diffusion are other processes... but they're not the principle that a diffractive optic operates on''. This type of hallucination is particularly concerning because it demonstrates internal coherence—the answer sounds technically sophisticated and relates to optics—while being substantively wrong about the core mechanism.}\looseness=-1

\subsubsection{\add{Incomplete Answers}}
\add{Our \textbf{Phase 1} analysis showed that \textit{Incomplete Answers}} were common, and manifested as both \textit{major omissions} and subtle omissions such as \textit{lacking details} that could impact a reader's understanding. We distinguished \textit{Incorrect Answers} from \textit{Incomplete Answers} since the latter can still result in correct high-level answers. An answer can be:
\begin{itemize}
    \item correct and complete, where it is faithful with sufficient information coverage;
    \item correct but incomplete, where it is true at a high level, but missing key details required for accurate interpretation or reproducibility, e.g., missing methodological choices or failing to surface table data;
    \item incorrect but complete, where it covers each required component of the answer but includes factual errors that make it impossible to use the output;
    \item incorrect and incomplete, where it misses essential information and introduces false claims, e.g., generic comparisons between works that also misattribute citations.
\end{itemize}
This distinction matters analytically and practically. Collapsing the categories would hide different failure modes, such as precision errors (incorrectness) versus recall errors (incompleteness). Practically, the way an expert or model developer might want to respond to or prevent the error would be different depending on whether it is an error of omission or incorrect. Incorrectness calls for verification and correction, which can be approached by auditing citations, checking details like numbers and equations, or scaffolding generation with more sophisticated information systems. Incompleteness can be addressed using prompting to drill down into details, or more sophisticated information extraction techniques to ensure table/figure linkage, enumeration of required items, and inclusion of method specifics. When answers were both incomplete and incorrect we applied both codes. 

\textit{Major omissions} ranged from omissions of large-scale content to omissions that left a response seemingly plausible but unusable in practice. Structural omissions included \textit{ignored section}, \textit{missed main point}, and \textit{ignored whole paper} in multi-document contexts. For example, when asked ``What are the main limitations of [\textsc{Approach}] when modeling a large number of [\textsc{Data Type}]?,'' the system discussed computational efficiency and complexity but completely missed the paper's central point about exponential growth in [\textsc{Data Type}] combinations. This omission hurts the usefulness of the answer, because the exponential scaling issue is the main limitation that drives all other computational concerns.\looseness=-1

Other \textit{major omissions} reflected missing components required for accurate interpretation or traceability, including \textit{incomplete definition}, \textit{incomplete references}, and \textit{chronological gaps}. For example, when asked ``How was the microphone performance characterized prior to deployment?'', the system correctly identified the four measurement types: ``measurements of absolute microphone sensitivity, frequency response, total harmonic distortion, and noise floor.'' However, it omitted crucial specifications present in the source: the sensitivity measurement specified exact calibration parameters (frequency and sound pressure level), the frequency response identified the specific calibration equipment used, and the noise floor measurement detailed the required testing environment and procedure. Our domain experts noted that the characterization of performance would be inaccurate without this specific information---the sensitivity of the performance measures differs based on these specifications. Any knowledge one would gain from simply focusing on the former would be misleading. In this way, subtle omissions could be just as harmful as major omissions by appearing to be sufficiently informative. 

Relatedly, we coded \textit{lacking details} when high-level concepts were present but lower-level procedural or parameter information was missing, and we coded \textit{multimedia comprehension} when the system failed to connect tables/figures with captions and in-text references.\looseness=-1

\add{In \textbf{Phase 2}, at least one sub-type of \textit{Incomplete Answers} came up in all 10 experts' unprimed evaluation.} Several experts described omissions that aligned with \textit{major omissions}, including \textit{missed main point}, \textit{ignored section}, and, in multi-document synthesis-based QA, \textit{ignored whole paper} \add{(see column Part 1 in Table \ref{tab:schema-2})}.  \add{For example, E10 noted ``the whole kind of bigger 10,000 foot view purpose was missed several times,'' while E8 observed responses that ``basically reworded what was said in the paper correctly, but did not answer the question.'' E7 found that the system listed input categories correctly but missed specific technical formats that distinguished the approach, reflecting \textit{lacking details} even when the higher-level framing was accurate.} Across sessions, experts also identified \textit{incomplete definition} and \textit{incomplete references} when key interpretive or traceability information was missing, as well as \textit{chronological gaps} when timelines omitted required events. Missing larger sections of the paper or, in the multi-document contexts, ignoring one or more papers entirely came up as forms of incompleteness in both unprimed and primed parts of \textbf{Phase 2}. This category was considered especially worrisome by E6}\add{---experts worried that end-users \final{unfamiliar with the paper} would not know when an important section or paper would have been ignored unless they had closely read the paper on their own.} 

\subsubsection{\add{Question Interpretation}}
\add{In \textbf{Phase 1}, we described} \textit{Question Interpretation} errors as cases where the LLM either showed \textit{question misinterpretation}—misunderstanding an aspect of the prompt in a way that led to an incorrect answer—or \textit{question redirection}, where it addressed a related but different question.  For example, when asked ``How were the differences between real-world and simulation data outputs reconciled?,'' the system discussed improving simulation fidelity rather than the actual data normalization methods used for comparison. We coded this as \textit{question misinterpretation} because the model interpreted ``reconciling differences'' as ``how would you improve simulation fidelity'' rather than ``how would you transform the data for comparison.'' In contrast, when asked ``What are the most influential sources?,'' intended to mean ``which prior works were most relied upon by the author,'' the model redirected into unrelated territory about various unspecified sources from other disciplines, reflecting \textit{question redirection} rather than an attempt to answer the original citation-focused question.

Not all interpretation differences counted as errors; for instance, when asked, ``How did the qualitative findings complement or contradict the quantitative results?,'' the model framed its answer in terms of triangulation and synthesis (how interviews contextualized or added nuance to survey trends) rather than itemized one-to-one agreement. We judged this an acceptable alternate reading of the prompt, even though that particular response also contained a separate referencing hallucination. In such cases, we attribute the mistake to the appropriate downstream category (e.g., \textit{citation information} or \textit{document structure} hallucinations such as invented table numbers) while treating the interpretation itself as reasonable.\looseness=-1

In \textbf{Phase 2}, the \textit{question misinterpretation} sub-type was brought up by almost all (8 out of 10) of the experts. \textit{Question Interpretation} was therefore frequently noted even unprimed in \textbf{Phase 2} by 9 out of 10 experts (see column Part 1 in Table \ref{tab:schema-2}). \add{Several experts observed that correct interpretation often required domain knowledge that the system lacked. E3 asked about benefits during ``nominal operation'' for a safety system that only engages during crash situations: ``The disconnect for the AI is that it doesn't understand that nominal operations—a crash situation cannot be considered a nominal operation... you'd have to have that background knowledge that those two are mutually exclusive.'' This illustrates how \textit{question misinterpretation} can cascade from gaps in domain understanding rather than simple parsing failures.

\textit{Question redirection}—where the system answers a related but different question—came up less frequently, but with high misleading potential. E10 asked how diffractive optics are made but received information about their properties and applications: ``Most everything here is true, but none of it is how they're made... none of it's wrong. But it's not the question that was asked.'' Similarly, E9 asked about experimental results but received design goals: ``It more answered the question of how was the actuator designed... it's accurately stating the design goal, it's just not accurately answering about the results.'' These cases are particularly problematic because the responses appear substantive and relevant, potentially masking the fact that the actual question went unanswered.}

\subsubsection{\add{Synthesis Issues}}
\add{In \textbf{Phase 1},} \textit{Synthesis Issues} emerged primarily in multi-document contexts, where the system failed to integrate information across sources in a coherent way. For example, when asked to trace the evolution of a topic of research across multiple papers, the system produced a \textit{disjointed response}, discussing each paper in isolation rather than synthesizing a chronological narrative. The model occasionally struggled to synthesize temporal information between papers or even across multiple paragraphs (see Figure \ref{fig:example}). The distinction between \textit{chronological confusion} (categorized under \textit{Synthesis Issues}) and \textit{chronological gaps} (categorized under \textit{Incomplete Answers}) reflects different types of temporal failures. \textit{Chronological gaps} occur when the system omits key events or developments from a timeline; e.g., when describing the evaluation of a device, the LLM mentioned ecological tests but omitted the primary internal tests that preceded them. This is an error of omission. In contrast, \textit{chronological confusion} represents a failure to correctly understand temporal relationships between events, such as presenting events out of sequence or misunderstanding which developments preceded others. 

Other instances of \textit{Synthesis Issues} include \textit{document linking failure}, where the system could not reliably connect a method, result, or claim to the correct document in a multi-document setting. In these cases, the response treated the papers as an undifferentiated pool of evidence, making it difficult to tell which document supported which statement. A related but more specific failure mode was \textit{source confusion}, where the system would mix findings from different papers when discussing methodology, attributing techniques described in one paper to experiments conducted in another. For example, when the system stated ``a central theme in all three papers'' when analyzing four papers, yet referenced content that only appeared in the fourth paper. This internal inconsistency suggests the model struggles with document tracking. These \textit{source confusion} errors are particularly problematic in scholarly contexts where proper attribution is critical for understanding research contributions. It is important to distinguish \textit{source confusion} from \textit{reference confusion} (a correctness error). Reference confusion occurs when the system incorrectly cites specific content, as when asked about modeling approaches across three papers and the system consistently mismatched paper years with their content. An expert noted the system seemed to be ``struggling to correctly keep track of which citation is which.'' While \textit{reference confusion} involves incorrect citations within a response, \textit{document linking failure} and \textit{source confusion} concern cross-document synthesis: the former reflects an inability to maintain stable document-to-claim mappings, while the latter reflects explicit misattribution across papers.\looseness=-1

\add{In \textbf{Phase 2}, most experts recognized some sub-type of \textit{Synthesis Issues} in their unprimed evaluation, with \textit{disjointed response} across multi-document questions being the most common sub-type.} They noted that the system responded without ``making the connection to the bigger picture'' (E8). \add{An important novel observation emerged in this phase that impacted \textit{Synthesis Issues} in particular}: while our authorial team's domain experts wrote questions within the closed-QA paradigm, the validation study experts sometimes posed questions that assumed open-domain capabilities. Despite our system's explicit instructions for the LLM to limit itself to provided documents and to acknowledge information gaps, the model would attempt to answer these questions by drawing on its pretrained knowledge. \add{This exacerbated the feeling of ``broad answers'' (E6) in response to specific synthesis requests.} 

\subsubsection{\add{Formatting Issues}}
\finaldelete{\textbf{Formatting issues} such as} \textit{Verbosity}, \textit{notation errors}, and \textit{inconsistent referencing} were grouped together. \add{While less severe than factual errors, they affect usability and can obscure important content. 

From \textbf{Phase 1}, \textit{verbosity} emerged as a frequent concern. When asked ``What is the main technical proposition of the paper?''—a focused, simple question—the system provided an extended response covering justifications, impact, and tangential context. The expert noted: ``This is a little bit more wordy than I would have liked... LLMs tend to be a little verbose. And unnecessarily so, because here we have a clear focus. It's a simple question. What is the main technical proposition. And it goes to justifications and impact and stuff like that. I didn't ask for that''. This pattern of over-generation, where the system provides unsolicited elaboration, was common across expert evaluations.
We also observed \textit{notation errors} where the system used incorrect mathematical symbols or variable names. In one case, an answer used $p_x$ where the paper used $p_a(x,t)$. The surrounding text was broadly accurate, but the notation errors made the result formally incorrect for anyone attempting to use or verify the equations.} We additionally coded \textit{language issues} when wording or grammar reduced clarity or precision, and \textit{inconsistent referencing} when the response used undefined or shifting referents that made it difficult to trace what a label or claim referred to.

\add{In \textbf{Phase 2},  formatting errors such as \textit{verbosity} and \textit{language issues} were cited often.} One example of formatting that was particularly confusing was when the answer referred to points A and B, but did not define what these were (E9), reflecting \textit{inconsistent referencing}. \add{Several experts noted that verbose responses made it harder to locate the relevant information, with E8 describing outputs that lacked focus on ``the bigger picture.'' While formatting issues alone may not invalidate an answer, they compound other errors by burying correct information in excessive text or introducing ambiguity through inconsistent notation and undefined references.}

\subsubsection{\add{System Failures}}
We also identified failures that are distinct from content errors because they capture technical limitations specific to the model that generated the outputs rather than problems with how the LLM processes and reasons about information. \add{Some of these limitations stem from the architectural constraints described in Section~\ref{sec:system}---the use of a smaller open-source model on local infrastructure without external API calls necessarily involves tradeoffs in retrieval capability and context integration.}

\add{In \textbf{Phase 1}, \textit{retrieval failures} manifested most commonly as the system claiming information was unavailable when it was explicitly present. When asked about microphone architecture, the system responded that the type of microphone architecture developed for the field-deployable phased array is not explicitly stated in the provided context---yet the entire purpose of the paper was to explain that architecture. Similarly, when asked about differences between flap configurations, the system claimed the specific differences in noise characteristics between these configurations are not explicitly stated, when in fact ``that was the entire point of the paper.'' These failures suggest the system could retrieve surface-level content but failed to recognize when retrieved content directly answered the question posed.} We also observed \textit{data structure issues}, where artifacts of the system's internal organization (e.g., missing or misaligned fields or document boundaries in the retrieved context) leaked into outputs and constrained what the model could faithfully report.\looseness=-1

\add{In \textbf{Phase 2},} experts that identified system failures frequently compared the system's performance to their own experiences with larger models. 
It helped that they were aware that not all models have multimodal capabilities, and they developed and refined mental models of the system's capabilities throughout the session. \add{Some \textit{LLM limitations} appeared as problems with multimedia capabilities. Note that these differ from the multimedia interpretation errors categorized under \textit{Incorrect Answer}. 

For example, E9 identified a case where the system claimed the ``specific numerical results are not provided in the paper'' when ``they are in a table on page 8... the data in the table shows that centralized designs provided more sound power reduction than decentralized designs.'' This suggests tabular data may not have been fully accessible to the retrieval system, reflecting \textit{retrieval failures}. Other experts identified similar issues with figures (E4) and appendices (E10). }\add{Beyond retrieval, experts noted failures in contextual integration---connecting information across document sections or with domain knowledge. E8 reflected: ``It's not quite understanding the bigger picture... it only understands this little paper and only that at a very remedial level. It's not making the connection to the bigger picture.'' E6 offered a complementary observation: ``The difference between a human and a large language model sometimes is context... there are things we take for granted that we just kind of know intuitively because we've spent our whole lives navigating the world.''}
\add{These system-level failures proved particularly difficult for experts to diagnose during unprimed evaluation, as they required comparing system outputs against known document contents. The structured inventory \final{appeared to} help experts systematically check for missing content types---tables, figures, sections, and cross-document information---that might otherwise go unnoticed.}\looseness=-1

\subsection{Expert Evaluation Patterns}\label{sec:expert-eval-patterns}
Our qualitative analysis also uncovered patterns in how domain experts approach the evaluation of LLMs in scholarly contexts. These patterns offer further evidence showing the \finaldelete{unique} \final{distinctive} approach taken by domain experts compared to automated or model developer-driven benchmarking of LLMs. 

\subsubsection{Systematic Assessment Strategies} 
Experts consistently used deliberate testing strategies to evaluate specific model capabilities. For example, they intentionally designed questions to probe the model at multiple levels, from basic accuracy to higher-order capabilities such as synthesis and reasoning (E4). E8 similarly ``tried to make a part of it easy and a part of it really hard... I was swinging for the fences to see if it can understand''. E6 wrote questions that were intentionally misleading to test whether the system would show confirmation bias. 

In addition to question design, all experts demonstrated rigorous verification practices, by cross-referencing the system's outputs with the papers in question. For example, E4 checked if information comes from the main text versus figures, and referred to the paper to see if the system referenced the correct sections. Experts were exacting in their evaluation of technical precision, focusing on numeric details, table and figure comprehension, and mathematical understanding. However, one area requiring precision that was not investigated by the majority of the experts was accuracy in citations and references. When the system included full citations in its outputs, only 1 out of 10 experts preemptively checked the correctness of the citations (others did not systematically check citations, but discovered incorrect citations incidentally). We hypothesize that this oversight likely reflects both unfamiliarity with citation hallucination as a common LLM failure mode and the practical burden of verifying bibliographic details during time-limited evaluation sessions.

Beyond these systematic processes, experts also drew on their prior experiences with LLM technology, especially commercial chatbots such as ChatGPT and Claude. They relied on this knowledge to qualitatively benchmark our system's performance against larger commercial models (E1, E2, E3, E6, E8, E10) and understood common LLM behavioral patterns: ``You know, it's because it wants to please... they start apologizing profusely'' (E2). These mental models evolved throughout the session as experts calibrated their expectations to our system's specific capabilities and limitations. 

\subsubsection{Value-Based Assessment} 
Experts prioritized certain system behaviors and values, and disliked others. One common value was honesty about system limitations, as most experts preferred acknowledgment of constraints over attempted answers that may be unreliable: ``A good answer would be I don't have enough information to answer that question. They [LLMs] find the most statistically probable next word... and they provide it instead of saying I don't know enough'' (E2). They liked the transparency specified by our system prompt, noting ``I really appreciate that it was willing to admit not being able to read the figures and therefore it didn't attempt to make an estimate based on hallucinating numbers. I think it is a very virtuous response'' (E1). E10 was similarly ``pleasantly surprised that it didn't try and make stuff up.'' This emphasis on honesty stemmed from concerns about scientific integrity when systems fail to acknowledge limitations. Experts worried about believable confabulations that could mislead readers (E8), particularly the general public (E6). For these experts, the system's willingness to admit uncertainty was critical for maintaining scientific integrity.

\subsubsection{Research Applications} 
Expert evaluations were grounded in the practical utility of using LLM-based systems for real workflows. They performed context-based ecological assessment, considering what they would need from a scholarly QA system in their institutional roles (E1-2), with E1 wanting to know ``how well could it try and do my job for me of taking the analysis and being able to present it to a senior decision maker.'' To that end, experts criticized the system's limited domain knowledge, that it was ``not quite understanding the bigger picture; it only understands this little paper and only that at a very remedial level.'' (E8). E3 noted a logical inconsistency caused by a lack of domain knowledge, since the system was unaware that two states were mutually exclusive. As such, E3 wished that the system could connect to an external service such as Google Scholar, because the ``scope of reference for answering the question is really limited to the paper itself''. E2 noted that ``any useful interaction that has to do with analysis of a paper must also have access to the world of papers.''

\subsubsection{Meta-Evaluation}
Experts frequently reflected on their own assessment practices, and were open to revising initial judgments. This occurred when they believed that their question formulation might have influenced the system's responses, leading them to deliberate on whether the system's interpretation was valid despite not aligning with their initial expectation. Their careful consideration of context and meta-evaluation is reflected in E1's comment: ``it ended up answering the question differently than I was expecting, I think. Maybe in part because of the way I asked it.''

\subsection{Question Characteristics}
\add{Finally, we considered how the type of question asked may influence the errors that were surfaced. To understand this, we categorized the 120 expert-written questions from \textbf{Phase 2} based on interview data rather than static question text, incorporating participants' narratives about their intent. We then applied these types to the original 68 questions from \textbf{Phase 1} as well. Our thematic analysis of questions yielded 11 categories (see Appendix~\ref{app:question_type_descr} for full descriptions). For each category below, the number in parentheses indicates the number of questions (out of 188) belonging to that category:}
\add{
\begin{itemize}
    \item Critical Evaluation \& Validation (29): Probed underlying assumptions, assessed potential biases, and evaluated whether research choices were adequately justified.
    \item Methodological Inquiry \& Improvement (28): Compared different approaches to identify best practices and explored how methods could be transferred or refined.
    \item Meta-Analysis \& Contextualization (25): Explored institutional motivations, synthesized insights across multiple papers, and addressed meta-aspects of the research process.
    \item Application \& Practical Implications (24): Examined how findings translate to real-world uses and what tradeoffs implementation might require.
    \item Technical Details \& Specifications (24): Requested specific experimental parameters, equations, or system configurations needed for replication.
    \item Definitions \& Concepts (20): Sought to clarify foundational ideas and ensure precise interpretation of key terms.
    \item Binary/Factual Verification (9): Required confirming or denying discrete facts rather than elaborating or interpreting.
    \item Procedural Information (9): Probed how studies were conducted, focusing on sequences of study design.
    \item Future Directions \& Extrapolation (8): Looked beyond current findings to identify next steps and predict future research trajectories.
    \item External Context (7): Required knowledge beyond the paper itself, such as current practices, external standards, or developments since publication.
    \item Numerical Analysis \& Derivation (5): Asked the system to perform calculations or mathematical reasoning with data from the paper.
\end{itemize}
}
\add{This distribution reflects the diversity of scholarly information needs. Domain experts did not limit themselves to simple factual queries; instead, they naturally posed questions requiring interpretation, critique, and synthesis.}

\begin{figure*}[t]
        \includegraphics[width=\linewidth]{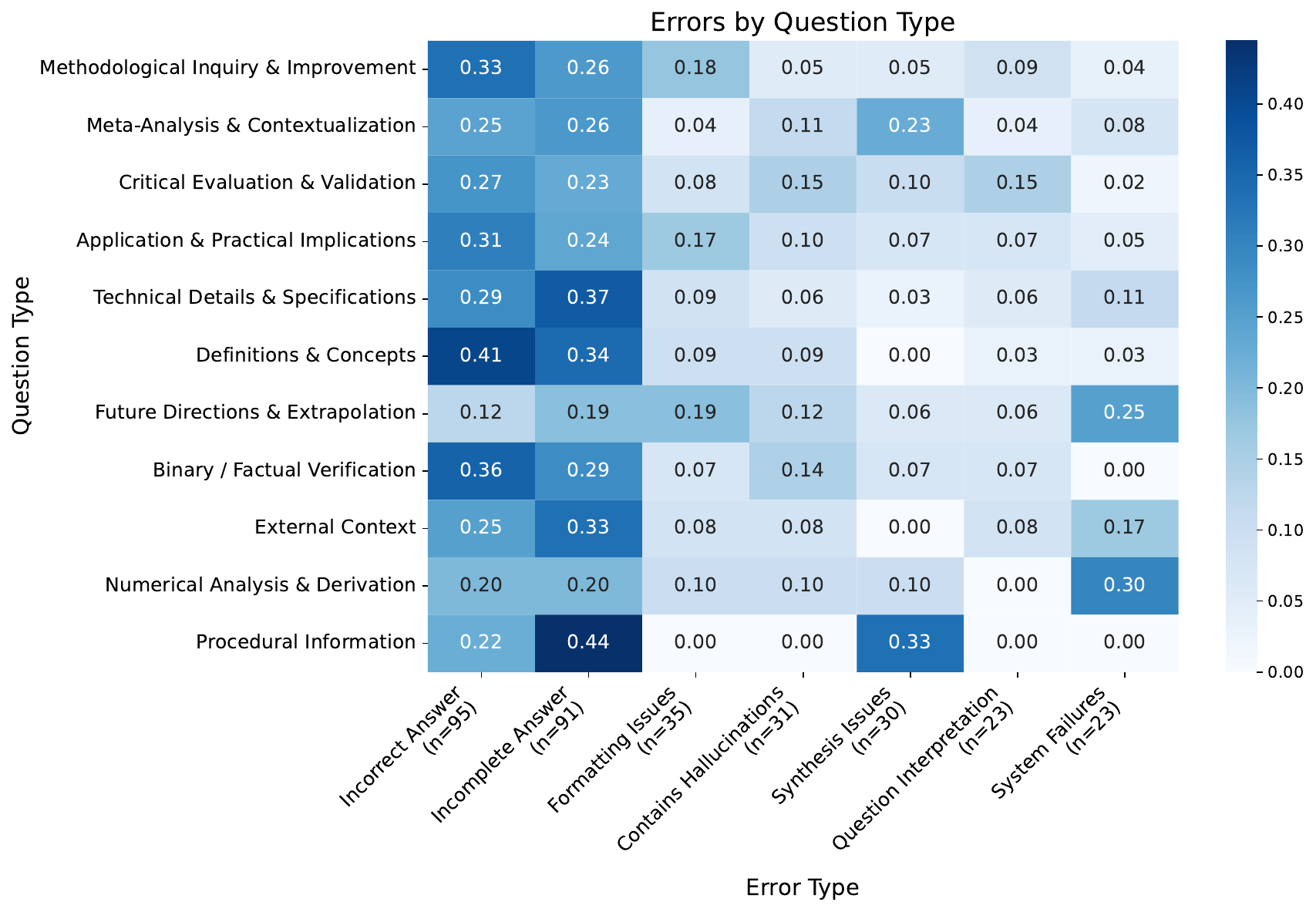}
        \vspace{-1.5pc}
        \Description{A heatmap titled ``Errors by Question Type'' displaying the proportional distribution of seven error types across eleven question categories. The y-axis is labeled ``Question Type'' and the x-axis ``Error Type.'' Cell shading uses a sequential blue color scale shown in a vertical legend on the right, ranging from white at 0.00 to dark navy at values above 0.40. Darker cells indicate that a given error type accounts for a larger share of errors for that question category. Each cell contains its exact numeric proportion.

The seven columns, from left to right, are: Incorrect Answer (n=95), Incomplete Answer (n=91), Formatting Issues (n=35), Contains Hallucinations (n=31), Synthesis Issues (n=30), Question Interpretation (n=23), and System Failures (n=23). The eleven rows, from top to bottom, are: Methodological Inquiry and Improvement, Meta-Analysis and Contextualization, Critical Evaluation and Validation, Application and Practical Implications, Technical Details and Specifications, Definitions and Concepts, Future Directions and Extrapolation, Binary / Factual Verification, External Context, Numerical Analysis and Derivation, and Procedural Information.

The dominant visual pattern is a dark band spanning the two leftmost columns, Incorrect Answer and Incomplete Answer, which together account for roughly 50 to 75 percent of errors across nearly all question types. Shading fades progressively lighter toward the right-hand columns. The single darkest cell in the heatmap is Procedural Information crossed with Incomplete Answer at 0.44, rendered in deep navy. Definitions and Concepts crossed with Incorrect Answer at 0.41 is comparably dark. Several cells break the otherwise pale right side of the heatmap: Procedural Information shows 0.33 for Synthesis Issues, Meta-Analysis and Contextualization shows 0.23 for Synthesis Issues, Numerical Analysis and Derivation shows 0.30 for System Failures, and Future Directions and Extrapolation shows 0.25 for System Failures---all rendered in darker blue than their neighbors in those columns. Multiple cells are 0.00 and appear white, including four of seven error types for Procedural Information (Formatting Issues, Contains Hallucinations, Question Interpretation, and System Failures), Synthesis Issues for both Definitions and Concepts and External Context, Question Interpretation for Numerical Analysis and Derivation, and System Failures for Binary / Factual Verification.}
        \caption{Distribution of error types across question categories. Each row is normalized to sum to 1, showing the proportion of errors within each question type. Column labels indicate the total occurrences of each error type across all questions ($
n$). Question types are sorted by total error frequency (descending); error types are sorted by total occurrences (descending). Darker cells indicate that a given error type accounts for a larger share of errors for that question category.}
    \vspace{-1pc}
    \label{fig:errors-by-qtype}
\end{figure*}

\add{\subsubsection{Error Patterns by Question Type}

Figure~\ref{fig:errors-by-qtype} shows how error types are distributed across the 11 question types. The four question types with the largest absolute number of error codes \finaldelete{are} \final{in our sample are} \textit{Methodological Inquiry \& Improvement}, \textit{Critical Evaluation \& Validation}, \textit{Application \& Practical Implications}, and \textit{Meta-Analysis \& Contextualization}. We view these as higher-order question types that require interpreting, evaluating, or applying research rather than recalling isolated facts. Therefore unsurprisingly, these higher-order questions account for a large share of observed errors, suggesting that RAG-based scholarly assistants struggle most when experts ask them to reason about study design, critique evidence, or extrapolate implications rather than answer pointwise factual prompts.\looseness=-1

\final{In our analysis,} across question types the most common failure modes when outputs were unsatisfactory were \textit{Incorrect Answer} and \textit{Incomplete Answer}. In the heatmap, these two error families dominate most rows: for example, Methodological questions are coded as Incorrect or Incomplete in 59\% of cases (0.33 + 0.26), and Critical Evaluation questions in 50\% (0.27 + 0.23). Even for relatively concrete categories such as \textit{Technical Details \& Specifications} and \textit{Binary / Factual Verification}, Incorrect and Incomplete together account for the majority of errors. 

The proportions also surface error types that are more prominent for particular question families. \textit{Future Directions \& Extrapolation} and \textit{Numerical Analysis \& Derivation} questions are \finaldelete{notably more likely to trigger} \final{associated with} \textit{System Failures}: 25\% and 30\% of their errors, respectively, are coded as failures (e.g., \textit{LLM limitations} or \textit{retrieval failures}, compared to near-zero rates for several other question types. \textit{Procedural Information} questions have the highest rate of \textit{Incomplete Answer} (44\%) and elevated \textit{Synthesis Issues} (33\%) \final{in our analysis}. This suggests that even the seemingly straightforward ``how-to'' questions \final{in our data} \finaldelete{often} require integrating steps scattered across the paper, and the scholarly QA system frequently failed to synthesize them correctly.

\textit{Hallucinations} constitute a smaller but non-trivial set of errors overall. They are most prevalent in \textit{Critical Evaluation \& Validation} (15\% of errors), \textit{Binary / Factual Verification} (14\%), and \textit{Meta-Analysis \& Contextualization} (11\%), with additional contributions from \textit{Future Directions \& Extrapolation} and \textit{Numerical Analysis \& Derivation}. In these categories, the model sometimes invents citations, misstates prior work, or fabricates numerical values when attempting to provide broader explanations or cross-paper comparisons. This supports experts' intuition that hallucinations are especially likely when the model is asked to generalize or contextualize beyond directly retrievable statements.

Our question-type analysis further illustrates why RAG-style architectures are insufficient on their own. Even when experts ask detailed, well-scoped methodological or critical questions, the system most often fails by being simply wrong or incomplete, rather than refusing to answer. At the same time, future-oriented and numerically intensive questions are disproportionately associated with system failures, and procedural questions frequently expose synthesis issues. \textit{Hallucinations} are less frequent overall but cluster in integrative and verification-oriented questions, where the model is pressured to provide context, summarize prior work, or reconcile multiple findings. These patterns reinforce that evaluation and mitigation strategies must be differentiated by question type: improving retrieval coverage will not address higher-order reasoning failures, while detecting hallucinations and ungrounded extrapolations is particularly important for contextual and meta-analytic use cases.}

\section{Discussion}\label{sec:discussion}
We present an expert-centered schema for evaluating LLM errors for scholarly QA, developed and validated with domain experts, NLP specialists, and HCI researchers. Our schema represents fine-grained details of contextual use that reflect an expert perspective, both in the evaluation codes identified and in the assessment strategies and values represented in the process. \add{Having established these categories of errors, we now return to our motivation of contextual evaluation of LLMs for scholarly QA and discuss: automation potential and hybrid evaluation approaches; salient expert needs for assessing LLM reliability in scholarly contexts; and the generalizability of our schema itself.}

\add{\subsection{Implications for Evaluation Approaches for Scholarly QA}}
\add{Our schema \final{may} offer a foundation for combining automated methods with expert oversight by \finaldelete{identifying} \final{suggesting} which error types are amenable to automation. For example, citation hallucinations (3.g), document structure hallucinations (3.f), retrieval failures (19), and notation errors (15) could be partially automated through cross-referencing digital libraries, detecting claims of missing information when relevant passages exist, and comparing extracted symbols against source documents. In this way, automatic evaluations could serve as first-pass filters that flag potential issues for expert review. In contrast, hallucinations of \textit{technical content} (3.d), \textit{missed main point} (5.b), \textit{Synthesis Issues}, and \textit{question misinterpretation} (8) resist automation because they require contextual understanding of domain principles, central contributions, and \final{domain-expert} intent that current systems lack. These errors are particularly \finaldelete{dangerous} \final{concerning} because they appear plausible.

Our findings suggest hybrid workflows where automated pre-screening addresses evaluation fatigue while preserving expert judgment for contextual errors is an ideal path forward. LLM-as-judge approaches could use our schema's granularity to check for specific failure modes, though our finding that experts missed fabricated citations until guided by the inventory suggests automated judges may share similar blind spots. Finally, schema-informed benchmarks could separately evaluate hallucination resistance, synthesis capability, and retrieval reliability, with question-type-specific protocols allocating resources based on our finding that methodological and critical questions disproportionately surface \textit{Incorrect Answers} while numerical questions trigger more \textit{System Failures}.}

\subsection{Designing for Expert Needs in Scholarly AI Systems}\label{sec:discussion-design-implications}

\subsubsection{Multimodal Comprehension. }Experts consistently highlighted the importance of multimodal comprehension, including tables, figures, and equations, indicating that they require systems that can maintain the semantic relationships between different representations of knowledge. While newer models have better multimodal capabilities, we still need interpretable approaches for comprehension between information generated via artifacts (e.g., figures, tables) and the main text, to aid in efficient verification. This transparent multimodal comprehension requires not only information extraction techniques from the artifacts themselves, but also resolution of mentions in the main text, as well as mathematical reasoning capabilities. Making these connections visible and verifiable is important for supporting experts in tracing how the system interprets multimodal content. Future models might learn from assistive technology on this front~\cite{sorge2014towards}.

\subsubsection{Uncertainty Communication. }Transparent descriptions of uncertainty were one of the values requested by our experts (E1, E10) and have been found in recent work to help reduce over-reliance \cite{10.1145/3630106.3658941}. However, the system's approach to uncertainty communication needs improvement as a critical design feature (E2). This suggests opportunities for designing uncertainty indicators that align with disciplinary norms and expert mental models. This value directly contradicts the use of larger, commercial LLMs for scholarly QA, as the lack of faithful interpretability metrics makes these models opaque. Future work on mechanistic interpretability methods will be critical for this expertise-driven context.

\subsubsection{Personalized Evaluations and Adaptive Interfaces. }Our schema offers potential mechanisms for personalizing evaluation of scholarly QA outputs. Even in our small validation group, experts varied in which errors they identified, and were fatigued by the larger inventory-based evaluation. This suggests an opportunity for personalizing the presentation of a schema-based inventory based on longitudinal observations of \add{expert} behaviors. 
Improving expert identification of errors might also consequently reduce over-reliance on LLM outputs~\cite{afroogh2024trust,liao2023ai}. 

These personalization opportunities point to broader implications for human-AI collaboration in scholarly work. The errors we identified, particularly hallucinations and synthesis failures, directly threaten core scholarly values of accuracy, attribution, and intellectual rigor. Yet our experts demonstrated sophisticated strategies for working with imperfect AI outputs, suggesting that effective human-AI collaboration is possible with appropriate support. 
Adaptive interfaces that learn from individual evaluation patterns could highlight error types that particular experts tend to miss while streamlining checks they perform reliably, addressing the evaluation fatigue raised by multiple experts while maintaining thoroughness. \add{Such personalization could be implemented through progressive disclosure interfaces that initially present high-risk error categories specific to each \final{domain-expert}'s blind spots, with additional evaluation dimensions available on demand. This approach could also support skill development by gradually reducing scaffolding as experts internalize particular error-checking practices, ultimately fostering more independent critical evaluation of AI-generated content. }

\subsection{Schema Generalizability}
While our schema was developed with experts in STEM fields, we \finaldelete{believe} \final{hypothesize} that the high-level axial codes generalize to other scholarly domains given the consistent fundamental processes experts use to evaluate LLM responses. Our qualitative approach showed that experts engaged in a structured analysis of claims, evidence, and warrants when evaluating answers, which is a process fundamental to scholarly argumentation across disciplines~\cite{Toulmin_1958}. However, we recognize that evaluation criteria may vary across epistemological traditions.\looseness=-1

Per \citet{10.1145/1142405.1142422}'s framework on multiple interpretations in design and evaluation, we suggest that our schema's value is not in imposing uniform evaluation criteria, but rather in providing a structure that can accommodate ``multiple, potentially competing interpretations'' \cite[p. 1]{10.1145/1142405.1142422}. Abandoning the presumption that a specific, authoritative interpretation is necessary can more fully address the complexity of how different communities assess technological systems. Even within our relatively homogeneous sample of experts, we observed conflicting evaluation priorities. For example, while some experts wanted comprehensive technical detail in responses, others criticized the same level of detail as verbose and preferred high-level summaries. Some valued the system's willingness to admit uncertainty, while others found such admissions frustrating when they expected definitive answers. This variation reflects what \citet{10.1145/1142405.1142422} call ``interpretive flexibility,'' where the same system behavior can be legitimately interpreted as a strength or weakness depending on the evaluator's needs and values. As such, we see our schema as a set of guidelines for error categories to be mindful of across domains grounded in similar forms of argumentation, but recognize that the specific subsets of errors may need additions. 

The adaptation process itself could be valuable for communities to articulate their implicit quality criteria. For instance, humanities scholars might expand our hallucination categories to include items such as ``anachronistic interpretations'' or ``misrepresented historical contexts.'' Social scientists might add categories for statistical misinterpretation or inappropriate generalization from samples. Medical researchers would likely emphasize errors in evidence hierarchies and clinical significance. These domain-specific instantiations would maintain our framework's organizational logic while reflecting the \finaldelete{unique} \final{distinctive} epistemological commitments and verification practices of different scholarly communities.

\subsection{Ethical Considerations}
\label{sec:ethical-considerations}

First, we must acknowledge that running inference even with smaller models has environmental impacts that are important to ethical considerations. Development, testing, and deployment of our system resulted in 61 GPU hours on Tesla V100 GPUs. More broadly, this work raises important ethical considerations about the role of LLMs in scholarly research. As these systems become more prevalent in industry and academic workflows, there are legitimate concerns about their potential impact on research quality, scholarly understanding, and intellectual agency. Our findings \finaldelete{reveal} \final{suggest} that even domain experts can miss certain types of errors without structured evaluation frameworks, showing that there are risks of inappropriate reliance on these systems. Additionally, the tendency of LLMs to generate plausible-sounding but potentially incorrect information could particularly impact early-career researchers or those working across disciplines who may lack the domain expertise needed to identify subtle errors and omissions. With QA and search being increasingly routed to LLMs, we as a field ought to provide guidance on where this could go wrong both for model developers but also domain experts; one of our goals with this schema development was to highlight that we, as researchers, are not ready to relegate our scholarly search to LLM-driven processes.

\section{Limitations}
\label{sec:limitations-and-future-work}
Our study has several important limitations. Our sample size (N=10) was small, and included experts within science and engineering. The schema may need adaptation for fields where the relationship between claims and evidence is more interpretive, such as in the humanities where multiple valid interpretations of the same evidence may coexist. While the fundamental categories of our schema (such as hallucinations and synthesis problems) should transfer, the criteria for what constitutes a satisfactory answer may differ in fields where scholarly argument relies more heavily on rhetorical analysis or philosophical reasoning. Replicating this study with experts in the arts, humanities, or social sciences would likely identify additional patterns in both system performance and expert evaluation.\looseness=-1

The technical implementation of our system also presents limitations. We deliberately chose an open-source model and retrieval-augmented generation pipeline for transparency and reproducibility. This choice aligns with practical constraints some researchers may face, where smaller models that can run on local workstations may be necessary in settings with strict security requirements.
\add{Critically for our evaluation goals, this choice also provided the interpretability necessary to distinguish between different error origins. For example, whether failures arose from retrieval issues, model limitations, or semantic misunderstandings, which would be impossible to determine with large commercial systems.} Additionally, while larger models might offer improved performance, their substantial computational requirements and environmental impact make smaller models worthy of investigation. However, this means our findings may not fully generalize to more powerful proprietary models. Future studies should examine which of our error types can be sufficiently addressed with technical solutions, \add{and benchmark different models using our schema.}

We also found that the 34-question inventory based on our schema was fatiguing for the experts to apply, highlighting a broader challenge around determining which quality assurance responsibilities should be delegated to automated systems versus which require human expert oversight. Future work could explore whether more advanced models could reliably self-identify certain error types, allowing experts to focus their evaluation efforts on aspects that \finaldelete{significantly} \final{particularly} require human reasoning, domain knowledge, and contextual understanding. This research direction is particularly important for making expert evaluation more sustainable and scalable while maintaining high standards of scholarly integrity.\looseness=-1

\section{Conclusion}
\label{sec:conclusion}
\finaldelete{We show}\add{Our findings suggest} that evaluating LLMs for nuanced tasks like scholarly QA requires contextual, qualitative methodologies and expert collaborations. Through such collaboration, we developed a schema of evaluation criteria for LLM outputs for scholarly QA. Procedurally, we discovered \finaldelete{unique} \final{distinctive} assessment patterns applied by domain experts, and their mental models of system capabilities and limitations. \add{We present these expert-generated artifacts and procedural understanding of their methods as opportunities for identifying hybrid and personalized evaluation approaches for LLM use in scholarly QA.}


\begin{acks}
We would like to thank our reviewers for their helpful comments. We are grateful to Braxton VanGundy and his team at NASA for their guidance on PDF text extraction. We are also grateful to Matthew Zent, Nuredin Ali, and all of the faculty and students in the GroupLens research lab for their feedback and support. We also want to thank everyone who participated in our study.
\end{acks}

\bibliographystyle{ACM-Reference-Format}
\bibliography{000_references}

\appendix

\section{Prompt Formatting} \label{app:prompt}
The prompting strategy we used includes descriptions of the LLM's role, its capabilities, the format of the input data, the task it is to perform, and final instructions and reminders. Following is the full prompt text.

You are an advanced AI research assistant designed to help users with scholarly literature analysis and question answering. Your primary function is to provide accurate and insightful answers to questions based on one or more scholarly papers.

Here are your key capabilities:
\begin{enumerate}
    \item You can analyze individual papers or sets of related papers, understanding their content, methodology, findings, and implications.
    \item You can answer questions about specific papers or synthesize information across multiple papers. Your answers should be directly relevant to the question asked.
    \item Your audience is researchers and domain experts. Therefore, your responses should be highly detailed, specific, and technical. You can assume the user has a high level of scientific literacy.
    \item Always provide specific citations or references to support your answers. Indicate which paper and section you are drawing from. Use direct quotes when possible. Encapsulate direct quotes in quotation marks.
    \item If a question cannot be answered based solely on the provided content, clearly state this limitation. Do not invent or assume information not present in the given materials.
    \item Communicate using a neutral, academic tone. Present information objectively, but you may point out any conflicting findings or interpretations across papers when relevant.
    \item If a user's question is ambiguous or could be interpreted in multiple ways, ask for clarification to ensure you provide the most relevant and accurate information.
    Overall, your goal is to assist researchers in understanding and extracting valuable information from scholarly literature while maintaining high standards of academic integrity.
\end{enumerate}
You will receive:
\begin{enumerate}
    \item A question about one or more scholarly papers; and
    \item JSON-formatted context containing relevant sentences from the paper(s).
\end{enumerate}
The question will be surrounded by the following tags: <question> and </question>. The context will be surrounded by the following tags: <context> and </context>. Ensure that both the question and context are present before proceeding. If either is missing respond with an error message. The context will be provided in the following JSON format:

\begin{verbatim}
{
  paper title: {
    "citation": author, year,
    "sections": {
      section name: [
        {
          "sentence_id": id,
          "sentence_text": text
        }
      ]
    }
  }    
}    
\end{verbatim}
Multiple paper titles may be present for multi-paper questions. Multiple sections will be present when there are multiple sections containing relevant
sentences.

Your task is to follow these steps to formulate your answer:
\begin{enumerate}
    \item Carefully read the provided question and context.
    \item Identify the key elements of the question (e.g. specific concepts, comparisons, or analyses required).
    \item Scan through the provided context to locate relevant information.
    \item For each relevant piece of information:
    \begin{enumerate}
        \item Note the paper title, section, and specific sentences it comes from.
        \item Evaluate its relevance and reliability in answering the question.
    \end{enumerate}
    \item If the context doesn't contain sufficient information to answer the question fully, explicitly state this limitation.
    \item Synthesize the relevant information to formulate a comprehensive answer.
    \item Ensure your answer is directly based on the provided context, not on your general knowledge.
    \item Review your answer to ensure it fully addresses all aspects of the question.
\end{enumerate}

Before submitting your final answer, review it to ensure that it directly answers the question asked, it is based solely on the provided context, and it includes correct references to papers and paper sections.
\color{black}
\section{Open Codes}\label{app:open-codes}
Table \ref{tab:open-codes} shows the open codes discovered by the model developer and the domain experts respectively.
\begin{table}[h]
    \centering
    \begin{tabular}{p{\linewidth}} \toprule
        \textbf{Common Codes:} Incomplete, Lacks detail, Repetitive, Unknown word or improper wording/spelling, Cut off/truncated, Misinterpreted the question, LLM refused to answer stating information not provided, Crucial/key information missing, irrelevant/superfluous, Lack of synthesis, Hallucinated citation, Source confusion; mixes up references and their content, Different interpretation of question, Hallucinated page numbers, Ignored paper in multidocument context, Mixes up papers and their content in multidocument context, Misleading, chronological misunderstanding/out of order, Inaccurate, Incorrect \\ \midrule
        \textbf{Unique to Model Developer:} Ignored whole section, Referenced wrong equation or math symbols, Missed main point, Misunderstood main point, No issues, Incomplete references, Minor omission, Missed information from table, Incomplete definition, Verbose, Hallucinated table, Mixes up information within document, Missed figure, Sentence IDs are irrelevant \\ \midrule
        \textbf{Unique to Domain Experts:} Improper referencing, Incorrect notation, Misleading due to ignoring large sections of paper, Misleading due to minor omission, Misleading in important way, Not comprehensive, Partially incorrect, Inconsistent referencing style, Vague/lacks specificity, Inconsistent with itself, Complete, Accurate, Thorough, All papers included in answer\looseness=-1\\
        \bottomrule
    \end{tabular}
    \caption{Open codes from the initial analysis, showing common codes found by both domain experts and model developers, and codes unique to each group. These reflect different priorities in evaluating LLM like 'Referenced wrong equation' and 'Missed information from table'.\looseness=-1}
    \label{tab:open-codes}
\end{table}
\section{Examples}\label{sec:examples}
\subsection{Hallucination of Technical Content}
\paragraph{Question:}What is a diffractive optic?
\paragraph{Answer:}A diffractive optic is an optical element that modifies the direction, phase, amplitude, or polarization of light through processes such as reflection, refraction, absorption, and diffusion...
\paragraph{Feedback:}The diffractive optics work based on the principle of diffraction as in what's in their name. Reflection, refraction, absorption and diffusion are other processes...but they're not the principle that a diffractive optic operates on.
\subsection{Synthesis Issues: Disjointed Answer}
\paragraph{Question:}What is the main benefit of using a clustering algorithm in the algorithm described in this paper?
\paragraph{Answer:}The main benefit of using a clustering algorithm in the [``\textsc{AlgorithmName}''] algorithm described in this paper is its ability to identify similarities between position solutions within a set of combinations of satellites. By grouping satellite combinations with similar position solutions, the algorithm can identify anomalies that deviate from these established clusters, such as outliers which may represent erroneous satellite measurements or unique error patterns. This approach enhances the detection of problematic satellite measurements and contributes to improved accuracy and reliability in navigation and positioning systems. 

Additionally, the clustering algorithm is used to determine the densest collection of position solutions, which helps in selecting the combination of satellites closest to the centroid of the cluster as the algorithm's solution. This method ensures a minimum number of valid position solutions are obtained and that the clustering algorithm can work effectively in reducing the state space of possible solutions by determining which satellite measurements are predicted to be NLOS and removing them prior to running the clustering algorithm.
\paragraph{Feedback:}... I am a little bit confused on the second paragraph... for example that first line on the second paragraph.. is true. The second part, there's no correlation to the first part. After the collection of position solutions, that part is true. That was the idea, but this which helps in selecting the combination of satellites closer to the center of the cluster as the algorithms. That connection is kind of off. They're still related, but it's like I wouldn't put it this way...
\subsection{Lacking Connections Between Information Formats}
\paragraph{Question:}Give an example of a regulatory statement which demonstrates compliance to a regulatory requirement
\paragraph{Answer:}An example of a direct regulatory statement can be found in section [\textsc{SectionNumber}]: ``The structure must be able to support limit loads without detrimental permanent deformation.'' This requirement imposes a specific structural limitation on the aircraft, and it is an explicit demonstration of compliance with a regulatory requirement for subsonic transport category aircraft.
\paragraph{Feedback:}So the regulatory statement is much longer than what's in the quotation marks here. And the way that it's supposed to be read is with a preceding text and there's even a figure that goes with it that shows how it's all connected.

\subsection{Incomplete Definition}
\paragraph{Question:}What constraints are used in this multi-point optimization?
\paragraph{Answer:}The constraints used in this multi-point optimization are the thrust, rolling moment, and pitching moment.
\paragraph{Feedback:}Again, it did not discusses constrain on design variables which is an important part of the optimization problem.

\subsection{Question Misinterpretation}
\paragraph{Question:}How were the differences between the real-world and the simulation data outputs reconciled to allow a legitimate comparison?
\paragraph{Answer:} The differences between the real-world and simulation data outputs were reconciled by adjusting and improving the simulated arrival capacity to match the real-world arrival capacity. Initially, the simulated arrival capacity peaked at [\textsc{CardinalNumber}] per hour, while the real-world [\textsc{Location}] advertises an arrival capacity of [\textsc{CardinalNumber}] per hour during peak operations in clear conditions. To resolve this shortfall, several improvements were made:

1. An assessment of traffic volume provided by the data set confirmed that the [\textsc{Dataset}] contained enough flights into [\textsc{Location}] to produce the required [\textsc{CardinalNumber}] per hour volume.
2. The configured routes were expanded to include modeling of tromboning, which significantly improved merging flexibility for the lower portion of the route and increased arrival capacity from [\textsc{CardinalNumber}] per hour for simulation.
3. Trajectory speed adjustments were added to maintain proper spacing between arriving aircraft, allowing simulated [\textsc{Location}] to consistently touch down aircraft with near minimal spacing during high volume operations, similar to real-world [\textsc{Location}].

These improvements allowed the simulated [\textsc{Location}] to demonstrate a good match to real [\textsc{Location}] when several critical flow techniques were captured for the modeled routes. Both the application of vectors and the trajectory speed adjustments used for the simulation route models are consistent with techniques used at [\textsc{Location}].
\paragraph{Feedback:}That wasn't really the point...we had to normalize them so that we could compare them...The data presentation that you typically get out of a simulation is very different style than the data presentation that you get from real world tools. I think that's a misinterpretation by the AI.

\subsection{Missed Main Point}
\paragraph{Question:}What are the main limitations of [\textsc{Approach}] when modeling a large number of satellites?
\paragraph{Answer:}The main limitations of [\textsc{Approach}] when modeling a large number of satellites are related to computational efficiency and complexity. Firstly, the naive or generalized ray tracing commonly employed is inefficient as it often repeats computational operations and does not directly compute the primary information that the urban satellite navigation visibility problem requires. Secondly, [\textsc{Approach}]'s formulation of the satellite visibility problem aims to improve computational efficiency; however, if a large number of satellites need to be modeled, other assumptions will likely be needed to reduce complexity again. Lastly, [\textsc{Approach}]'s output stage computes DOP for all possible satellite visibility combinations, which may become computationally intensive with a larger number of satellites.
\paragraph{Feedback:}So this one I would say I'm pretty dissatisfied... as we increase the number of satellites... the number of satellite combinations that are observed in this model grow exponentially... So this is the paragraph in the same page where we discussed that and that's what I was expecting it to touch on it.

\subsection{Misinterpreted Paper Content, Hallucinated Definition of Acronym}
\paragraph{Question:}I have a lot of higher frequency BVI noise in my computation.  How does the compact assumption influence the noise in the higher frequency?  Do I look accuracy in the BVI noise if I use compact thickness noise computation?
\paragraph{Answer:} The question asks about the influence of the compact assumption on higher frequency noise in the context of BVI (Boundary Value Problem) computation and whether accuracy can be improved in the BVI noise if compact thickness noise computation is used. 

The compact assumption is applied to the monopole term in [\textsc{Formulation}] for predicting rotorcraft noise. This approach relies on a formulation by Isom (Isom, 1997) and has been shown to be effective in understanding in-plane rotor noise and reducing computation time compared to noncompact forms. However, the compact assumption may not always accurately predict high frequency fall off rates and can introduce errors.

In Section IV of the paper, [\textsc{Citation}] explores the effect of blade shape on high frequency fall off rate and estimates error incurred when utilizing the compact assumption. The compact formulation assumes that all sound sources within the blade arrive at the observer at the same time, which can lead to reduced errors for observers located towards the rear of the rotor.

In Section [\textsc{SectionNumber}] presents a method for reducing error caused by the compact assumption. The proposed solution involves adding additional compact lines within the source volume that each account for a portion of the enclosed volume. This technique has been demonstrated to reduce errors incurred by the compact assumption.

To answer the question directly:

The compact assumption can influence higher frequency noise in BVI computation. While it can be effective in reducing computation time and understanding in-plane rotor noise, it may not always accurately predict high frequency fall off rates and can introduce errors. The paper by [\textsc{Citation}] discusses methods for estimating these errors and proposes a solution involving additional compact lines within the source volume to reduce errors. However, the question of whether accuracy can be improved in BVI noise if compact thickness noise computation is used remains unclear from the provided context.
\paragraph{Feedback:}So BVI is blade vortex interaction. So right off the the top it's not, it doesn't understand what that is...Furthermore, BVI is a dipole term, not a monopole term...So I think I mentioned that in the paper and it didn't pick it up here or I mentioned that that loading noise is is a dipole term, which is what BVI is.

\subsection{Verbosity}
\paragraph{Question:}What is the main technical proposition of the paper?
\paragraph{Answer:}The main technical proposition is the introduction of [\textsc{Framework}], a framework for systematic derivation of concepts for management and control of airspace systems. This approach is based on optimization formulations in terms of required system functions and states, aiming to enable ab initio design by resolving the lack of predictive modeling and traditional point design approach to development. The proposed methodology is intended to facilitate the daunting task of re-designing the air transportation system from scratch, addressing the challenges of transitioning from the present system to a fundamentally new one, and ensuring safety validation and certification of massively automated systems...
\paragraph{Feedback:}This is little bit more wordy than I would have liked... LLMs... tend to be a little verbose... And unnecessarily so, because here we have a clear focus. It's a simple question. What is the main technical proposition. And it goes to justifications and impact and stuff like that. I didn't ask for that.

\subsection{Superfluous Content}
\paragraph{Question:}Are there methods in one or more of these papers that can be applied to the problems outlined in the remainder of the papers?
\paragraph{Answer:} The question asks whether there are methods in one or more of these papers that can be applied to the problems outlined in the remainder of the papers. After analyzing the provided context, we can identify several methods that could potentially be applied to problems outlined in other papers.

In the "[\textsc{PaperTitle}]" paper, [\textsc{Citation}] proposes an approach called [\textsc{Framework}] for systematic development of concepts for management and control of airspace systems. This method is based on optimization formulations in terms of required system functions and states. The author argues that the incremental evolutionary approach to improving the air transportation system no longer suffices, and a new paradigm shift in design methods is needed. [\textsc{Framework}] could potentially be applied to problems outlined in other papers that deal with complex systems requiring systematic development and optimization.

In the "[\textsc{PaperTitle}]" paper, [\textsc{Citation}] present a method for quantifying uncertainty in computational models used in [\textsc{System}]. The authors use [\textsc{Approach}], which can estimate both the expected trajectory and its uncertainty. This method could potentially be applied to problems outlined in other papers that require [\textsc{Task}] in [\textsc{System}].

It is important to note that while these methods may have potential applications in other contexts, a thorough analysis of each specific problem would be necessary to determine whether these methods are indeed applicable and how they should be adapted to fit the specific requirements of each problem.
\paragraph{Feedback:}This is very generic... It was struggling just to find the words when it should have said ``I don't know''... It's just trying to please us and it's not doing a good job.
\section{Question Inventory}\label{app:inventory}
This section contains the inventory that was provided to the domain experts to structure their evaluation of LLM outputs in phase 2 of the schema validation study (see Table \ref{tab:inventory-1}). 
\begin{table*}[ht!]
    \centering
    \small
    \begin{tabular}{p{.03\textwidth}p{.81\textwidth}p{.08\textwidth}}\toprule
         \textbf{ID} & \textbf{Question} & \textbf{Schema \#} \\ \midrule 
         \multicolumn{3}{p{.92\textwidth}}{\textbf{Can you evaluate the answer’s completeness using the following questions?}}  \\ 
         Q1 &  Are there any major sections from the paper(s) that the system seemed to ignore that would have been necessary for a complete answer?  & 5.a \\
         Q2 & Are there any major topics or points from the paper(s) that the system seemed to ignore that would have been necessary for a complete answer?  & 5.b \\
         Q3 & In a multi-document setting, does the answer include information from all relevant papers mentioned in your question?  & 5.c \\
         Q4 & If the answer defines technical terms, are the definitions complete?  & 5.d \\
         Q5 & If the answer references figures, tables, or cited works, are the references complete? & 5.e \\
         Q6 & For topics involving a sequence of events, does the answer maintain a clear and complete timeline?  & 5.f \\
         Q7 & Does the answer provide sufficient supporting details for high-level concepts?  & 6 \\
         Q8 & If the answer references figures or tables, does it seem to leverage captions and in-text explanations to provide a thorough explanation of the referenced information?  & 7 \\ \midrule
        
         \multicolumn{3}{p{.92\textwidth}}{\textbf{Can you evaluate the answer’s correctness using the following questions?}}  \\
         Q9 & Are the main claims in the answer supported by the source materials?  & 1 \\
         Q10 & Are all specific details provided by the system correct?  & 2.a\\
         Q11 & Does the answer correctly interpret the meaning of the paper content?  & 2.b \\
         Q12 & Do any parts of the answer contradict each other?  & 2.c \\
         Q13 & Are references and the content they refer to consistent? For example, if the answer references Table 2, is it actually talking about Table 2, or a different table?  & 2.d \\
         Q14 & Does the answer stay focused on your question?  & 4 \\
         Q15 & Are there irrelevant citations or otherwise tangential or superfluous information?  & 4 \\
         Q16 & Are technical terms used correctly and consistently? Are acronyms correct?  & 3.a \\
         Q17 & Are referenced pages, papers, sections, tables, or figures real?  & 3.f \\
         Q18 & If the answer provides a citation, is all of the information in the citation correct? (Author names, year, publication venue, title, DOI, etc.)  & 3.g \\
         Q19 & If there is numerical information in the answer, is it correct?  & 3.b \\\midrule
         \multicolumn{3}{p{.92\textwidth}}{\textbf{Can you evaluate how well the system interpreted your question using the following questions?}}  \\ 
         Q20 &  Do you think the system interpreted what you were asking correctly?  & 8 \\
         Q21 & If your question had multiple parts, were all parts answered? & 8 \\
         Q22 & Did the system interpret all technical terms correctly? & 8 \\
         Q23 & Did the system answer the question you asked, or did it seem to answer a different question?  & 9 \\
         \midrule
        
         \multicolumn{3}{p{.92\textwidth}}{\textbf{Can you evaluate how well the system synthesized information in its answer using the following questions?}}  \\ 
         Q24 & Is the chronological development of ideas clear when relevant? & 11 \\
         Q25 & Is information presented in a logical flow?  & 12 \\
         Q26 & In a multi-document setting, does the answer make connections between related information from different papers? Is the information well-integrated, or does it feel like separate summaries?  & 12 \\
         Q27 & In a multi-document setting, is it clear which paper each claim or bit of information comes from? Does the answer attribute information to the correct paper?  & 13 \\ \midrule
        
         \multicolumn{3}{p{.92\textwidth}}{\textbf{Can you evaluate how well the answer is formatted using the following questions?}}  \\ 
         Q28 & Is the answer clear and concise?  & 14 \\
         Q29 & Is the technical notation used correctly and consistently? & 15 \\
         Q30 & Is the writing grammatically correct and well-structured?  & 16 \\
         Q31 & Are references and citations formatted consistently? & 17 \\ 
         \bottomrule
    \end{tabular}
    \caption{Question inventory for assessing LLM responses to scholarly questions. Column 3 shows which schema item the question maps to.}
    \label{tab:inventory-1}
\end{table*}

\section{\add{Question Type Descriptions}}\label{app:question_type_descr}
\add{
This section describes the different kinds of probing questions asked by domain experts. 

\textbf{Application \& Practical Implications.}
This theme describes questions that explore how research findings can be translated into real-world uses and what their practical consequences might be. They examine the potential applications and implementations of proposed methods or technologies, assessing their benefits, drawbacks, and performance characteristics in practical contexts. They investigate efficiency gains, resource utilization, and the tradeoffs inherent in different approaches, such as balancing noise reduction against propeller efficiency or comparing methods for multidisciplinary optimization. These questions help determine whether research advances are practically viable and what compromises might be necessary for implementation.

\textbf{Binary / Factual Verification.}
This theme includes short, verification-oriented questions that require the system to confirm or deny factual statements or identify discrete elements (e.g., parameters, components, or compliance indicators). They test the model’s ability to locate and verify facts within the document rather than to elaborate or interpret. These questions often take the form of yes/no checks or enumerations, such as identifying specific stereotypes, constraints, or design variables. Because they rely on factual accuracy rather than synthesis, they provide a useful baseline for evaluating model reliability on concrete information.

\textbf{Critical Evaluation \& Validation.}
This theme describes questions that critically examine the research's validity, rigor, and limitations. These questions probe the underlying assumptions and simplifications made in the research, assess potential biases or errors, and evaluate whether the authors' methodological choices and conclusions are adequately justified. They challenge the completeness and persuasiveness of the work by asking what was omitted and why, whether alternative approaches should have been used, and what additional evidence would strengthen the claims. These questions require deep domain expertise to recognize what should have been done differently and to identify subtle methodological weaknesses that may not be explicitly acknowledged in the paper.

\textbf{Definitions \& Concepts.}
This theme includes questions that seek to clarify foundational ideas, key terms, or central mechanisms introduced in the paper. These questions aim to establish conceptual understanding and ensure precise interpretation of the research’s core constructs, metrics, and mechanisms. They often ask what a particular concept means, how it is operationalized, or what fundamental processes underlie a system’s behavior. By grounding understanding in well-defined terms, these questions help ensure shared conceptual clarity across experts and systems.

\textbf{External Context.}
This theme describes questions that require knowledge beyond what is contained in the paper itself, drawing on broader understanding of the field, current practices, or general knowledge to properly contextualize the research. These questions explore how the research relates to the current state of practice in the field, compare proposed approaches with existing methods, or require bringing in external knowledge to expand on the paper's content. They may require understanding of external standards and regulations not fully explained in the paper, assess the research's real-world impact and reception, or ask about developments since publication. Some questions require adapting technical content for different audiences. These questions test the ability to situate the research within a broader knowledge landscape and understand its relationship to current practices, external standards, and real-world contexts that extend beyond the paper's content. 

\textbf{Future Directions \& Extrapolation.}
This theme describes questions that look beyond the current research to explore what comes next and where the field might be heading. These questions seek to identify logical next steps in the research trajectory, potential extensions of the work, and promising avenues for future investigation. They ask about follow-up studies that have already occurred or should occur, and attempt to extrapolate from current findings to predict future research questions or outcomes. These questions demonstrate forward-thinking engagement with the research and consideration of how the current work fits into a broader trajectory of scientific progress. 

\textbf{Meta-Analysis \& Contextualization.}
This theme describes questions that examine the broader context surrounding the research, looking beyond technical content to understand the scholarly and institutional landscape. These questions explore the motivations and driving forces behind the research, including institutional involvement and strategic decisions. They seek to synthesize insights across multiple papers to identify common themes, patterns, and connections that emerge from examining research collectively rather than in isolation. Additionally, they address meta-aspects of the research process itself, including editorial decisions, presentation strategies, and collaboration patterns. These questions demonstrate sophisticated engagement with research as part of a larger scholarly conversation and institutional context.

\textbf{Methodological Inquiry \& Improvement.}
This theme describes questions that examine the research methods themselves and explore ways to enhance, transfer, or better understand them. These questions compare different methodological approaches to identify best practices and optimal techniques for specific tasks, such as determining the most effective object detection algorithm or noise reduction method. They investigate how methods can be transferred across different contexts or problems, seek detailed explanations of how particular techniques or systems function, and propose improvements or expansions to the research methodology. These questions demonstrate engagement with the methodological rigor of the research and consideration of how the approaches could be refined, extended, or applied more broadly. 

\textbf{Numerical Analysis \& Derivation.}
This theme describes questions that require the system to perform quantitative analysis, calculations, or mathematical reasoning based on information in the paper, asking it to actively work with the data and equations presented to go beyond comprehension to perform its own computational tasks. Examples span from complex derivations requiring step-by-step mathematical work, to extracting specific values from data visualizations, to calculating computational requirements and performance improvements. These questions test the ability to manipulate numerical information, understand quantitative relationships, and make informed projections from data, requiring mathematical skills and domain understanding to complete successfully.

\textbf{Procedural Information.}
This theme describes questions that probe how a study was conducted—the processes, workflows, or analogies used in developing or validating the research. They focus on procedural reasoning rather than static facts, often asking how data were generated, how simulations or experiments were structured, or what methodological analogies underpinned a numerical model. These questions evaluate whether the model can reconstruct sequences of steps or methodological logic from the text, a skill central to understanding scientific process. 

\textbf{Technical Details \& Specifications.}
This theme captures questions that request specific technical, experimental, or computational details from the paper. These include inquiries about datasets, equations, parameter values, constraints, system configurations, and physical setups. They often require the model to extract fine-grained procedural or quantitative details that are crucial for replicating or extending the work, such as boundary conditions, optimization parameters, or sensor types. These questions test the model’s precision in locating and accurately reproducing detailed factual content embedded within the text. 

This question-type distribution shows diversity across retrieval, methodological, application, meta-analysis, etc., rather than focusing solely on one narrow task type.
}

\end{document}